\documentclass{llncs}

\usepackage[english]{babel}
\usepackage[utf8]{inputenc}
\usepackage{subcaption}

\usepackage{soul}
\usepackage{color}
\usepackage{slashbox}
\usepackage{mathtools}
\usepackage{amssymb}

\usepackage{graphicx}
\usepackage{epstopdf}
\usepackage{epsfig}
\graphicspath{{imgs/}{../imgs/}}

\usepackage[hidelinks]{hyperref}
\hypersetup{
  colorlinks = true,
  urlcolor   = blue,
  linkcolor  = blue,
  citecolor  = red
}

\title{
The Impact of Coevolution and Abstention on the Emergence of Cooperation
\thanks{The final version will be available at \textit{Studies in Computational Intelligence (SCI), Springer, 2017.}}
}

\author{Marcos Cardinot \and Colm O'Riordan \and Josephine Griffith}

\institute{Dept. of Information Technology, National University of Ireland, Galway, Ireland \\
\email{marcos.cardinot@nuigalway.ie}}

\index{Cardinot, Marcos}
\index{O'Riordan, Colm}
\index{Griffith, Josephine}

\begin{document}

\maketitle

\begin{abstract}

This paper explores the Coevolutionary Optional Prisoner's Dilemma (COPD) game,
which is a simple model to coevolve game strategy and link weights of agents
playing the Optional Prisoner's Dilemma game.
We consider a population of agents placed in a lattice grid with boundary conditions.
A number of Monte Carlo simulations are performed to investigate the impacts of
the COPD game on the emergence of cooperation.
Results show that the coevolutionary rules enable cooperators to survive and even
dominate, with the presence of abstainers in the population playing a key role
in the protection of cooperators against exploitation from defectors.
We observe that in adverse conditions such as when the initial population
of abstainers is too scarce/abundant, or when the temptation to defect is
very high, cooperation has no chance of emerging. However, when the simple coevolutionary
rules are applied, cooperators flourish.

\end{abstract}

\section{Introduction}
\label{sec:introduction}

Evolutionary game theory in spatial environments has attracted much interest
from researchers who seek to understand cooperative behaviour among rational
individuals in complex environments. Many models have considered the scenarios
where participant’s interactions are constrained by particular graph
topologies, such as lattices \cite{Hauert2002,Nowak1992}, small-world graphs
\cite{Chen2008,Fu2007}, scale-free graphs \cite{Szolnoki2016,Xia2015} and,
bipartite graphs \cite{Gomez2011}. It has been shown that the spatial
organisation of strategies on these topologies affects the evolution of
cooperation \cite{Cardinot2016sab}.

The Prisoner's Dilemma (PD) game remains one of the most studied games in
evolutionary game theory as it provides a simple and powerful framework to
illustrate the conflicts inherent in the formation of cooperation. In addition,
some extensions of the PD game, such as the Optional Prisoner's Dilemma (OPD)
game, have been studied in an effort to investigate how levels of cooperation
can be increased. In the OPD game, participants are afforded a third option ---
that of abstaining and not playing and thus obtaining the loner's payoff ($L$).
Incorporating this concept of abstention leads to a three-strategy game where
participants can choose to cooperate, defect or abstain from a game
interaction.

The vast majority of the spatial models in previous work have used static and
unweighted networks. However, in many social scenarios that we wish to model,
such as social networks and real biological networks, the number of
individuals, their connections and environment are often dynamic. Thus, recent
studies have also investigated the effects of evolutionary games played on
dynamically weighted networks
\cite{Huang2015,Wang2014,Cao2011,Szolnoki2009,Zimmermann2004} where it has
been shown that the coevolution of both networks and game strategies can play a
key role in resolving social dilemmas in a more realistic scenario.

In this paper we define and explore the Coevolutionary Optional Prisoner's
Dilemma (COPD) game, which is a simple coevolutionary spatial model where both
the game strategies and the link weights between agents evolve over time. In
this model, the interaction between agents is described by an OPD game.
Previous research on spatial games has shown that when the temptation to defect
is high, defection is the dominant strategy in most cases. We believe that the
combination of both optional games and coevolutionary rules can help in the
emergence of cooperation in a wider range of scenarios.

Thus, given the Coevolutionary Optional Prisoner's Dilemma game (i.e., an OPD
game in a spatial environment, where links between agents can be evolved), the
aims of the work are to understand the effect of varying the parameters $T$
(temptation to defect), $L$ (loner's payoff), $\Delta$ and $\delta$ for both
unbiased and biased environments.

By investigating the effect of these parameters, we aim to:
\begin{itemize}
    \item Compare the outcomes of the COPD game with other games.

    \item Explore the impact of the link update rules and its properties.

    \item Investigate the evolution of cooperation when abstainers are present
        in the population.

    \item Investigate how many abstainers would be necessary to guarantee
        robust cooperation.
\end{itemize}

The results show that cooperation emerges even in extremely adverse scenarios
where the temptation to defect is almost at its maximum. It can be observed
that the presence of the abstainers are fundamental in protecting cooperators
from invasion. In general, it is shown that, when the coevolutionary rules are
used, cooperators do much better, being also able to dominate the whole
population in many cases. Moreover, for some settings, we also observe
interesting phenomena of cyclic competition between the three strategies, in
which abstainers invade defectors, defectors invade cooperators and cooperators
invade abstainers.

The paper outline is as follows: Section~\ref{sec:related} presents a brief
overview of the previous work in both spatial evolutionary game theory with
dynamic networks and in the Optional Prisoner's Dilemma game.
Section~\ref{sec:methodology} gives an overview of the methodology employed,
outlining the Optional Prisoner's Dilemma payoff matrix, the coevolutionary
model used (Monte Carlo simulation), the strategy and link weight update rules,
and the parameter values that are varied in order to explore the effect of
coevolving both strategies and link weights.
Section~\ref{sec:results1} discusses the benefits of combining the concept of
abstention and coevolution.
Section~\ref{sec:results2} further explore the effect of using the COPD game
in an unbiased environment.
Section~\ref{sec:results3} investigates the robustness of cooperative behaviour
in a biased environment.
Finally, Section~\ref{sec:conclusion} summarizes the main conclusions and
outlines future work.

\section{Related Work}
\label{sec:related}

The use of coevolutionary rules constitute a new trend in evolutionary game
theory.  These rules were first introduced by Zimmermann et al.
\cite{Zimmermann2001}, who proposed a model in which agents can adapt their
neighbourhood during a dynamical evolution of game strategy and graph topology.
Their model uses computer simulations to implement two rules: firstly, agents
playing the Prisoner's Dilemma game update their strategy (cooperate or defect)
by imitating the strategy of an agent  in their neighbourhood with a higher
payoff; and secondly, the network is updated by allowing defectors to break
their connection with other defectors and replace the connection with a
connection to a new neighbour selected randomly from the whole network.
Results show that such an adaptation of the network is responsible for an
increase in cooperation.

In fact, as stated by Perc and Szolnoki \cite{Perc2010}, the spatial
coevolutionary game is a natural upgrade of the traditional spatial
evolutionary game initially proposed by Nowak and May \cite{Nowak1992}, who
considered static and unweighted networks in which each individual can interact
only with its immediate neighbours. In general, it has been shown that
coevolving the spatial structure can promote the emergence of cooperation in
many scenarios \cite{Wang2014,Cao2011}, but the understanding of cooperative
behaviour is still one of the central issues in evolutionary game theory.

Szolnoki and Perc \cite{Szolnoki2009} proposed a study of the impact of
coevolutionary rules on the spatial version of three different games, i.e., the
Prisoner's Dilemma, the Snow Drift and the Stag Hunt game. They introduce the
concept of a teaching activity, which quantifies the ability of each agent to
enforce its strategy on the opponent. It means that agents with higher teaching
activity are more likely to reproduce than those with a low teaching activity.
Differing from previous research \cite{Zimmermann2004,Zimmermann2001}, they
also consider coevolution affecting either only the defectors or only the
cooperators. They discuss that, in both cases and irrespective of the applied
game, their coevolutionary model is much more beneficial to the cooperators
than that of the traditional model.

Huang et al. \cite{Huang2015} present a new model for the coevolution of game
strategy and link weight. They consider a population of $100 \times 100$ agents
arranged on a regular lattice network which is evolved through a Monte Carlo
simulation.  An agent's interaction is described by the classical Prisoner's
Dilemma with a normalized payoff matrix. A new parameter, $\Delta/\delta$, is
defined as the link weight amplitude and is calculated as the ratio of
$\Delta/\delta$. They found that some values of $\Delta/\delta$ can provide the
best environment for the evolution of cooperation. They also found that their
coevolutionary model can promote cooperation efficiently even when the
temptation of defection is high.

In addition to investigations of the classical Prisoner's Dilemma on spatial
environments, some extensions of this game have also been explored as a means
to favour the emergence of cooperative behaviour. For instance, the Optional
Prisoner's Dilemma game, which introduces the concept of abstention, has been
studied since Batali and Kitcher \cite{Batali1995}. In their work, they
proposed the opt-out or ``loner's'' strategy in which agents could choose to
abstain from playing the game, as a third option, in order to avoid cooperating
with known defectors.  There have been a number of recent studies exploring
this type of game \cite{Xia2015,Ghang2015,Olejarz2015,Jeong2014,Hauert2008}.
Cardinot et al. \cite{Cardinot2016sab} discuss that, with the introduction of
abstainers, it is possible to observe new phenomena and, in a larger range of
scenarios, cooperators can be robust to invasion by defectors and can
dominate.

Although recent work has discussed the inclusion of optional games with
coevolutionary rules \cite{Cardinot2016ecta}, this still needs to be investigated
in a wider range of scenarios. Therefore, our work aims to combine both of
these trends in evolutionary game theory in order to identify favourable
configurations for the emergence of cooperation in adverse scenarios, where,
for example, the temptation to defect is very high or when the initial
population of abstainers is either very scarce or very abundant.

\section{Methodology}
\label{sec:methodology}

The goal of the experiments outlined in this section is to investigate the
environmental settings when coevolution of both strategy and link weights of
the Optional Prisoner's Dilemma on a weighted network takes place.

This section includes a complete description of the Optional Prisoner's
Dilemma (PD) game, the spatial environment and the coevolutionary rules for
both the strategy and link weights. Finally, we also outline the experimental
set-up.

In the classical version of the Prisoner's Dilemma, two agents can choose
either cooperation or defection. Hence, there are four payoffs associated with
each pairwise interaction between the two agents. In consonance with common
practice \cite{Huang2015,Nowak1992}, payoffs are characterized by the reward
for mutual cooperation ($R=1$), punishment for mutual defection ($P=0$),
sucker's payoff ($S=0$) and temptation to defect ($T=b$, where $1<b<2$).
Note that this parametrization refers to the weak version of the Prisoner's
Dilemma game, where $P$ can be equal to $S$ without destroying the nature of
the dilemma. In this way, the constraints $T > R > P \ge S$ maintain the dilemma.

The extended version of the PD game presented in this paper includes the
concept of abstention, in which agents can not only cooperate ($C$) or defect
($D$) but can also choose to abstain ($A$) from a game interaction, obtaining
the loner's payoff ($L=l$) which is awarded to both players if one or both
abstain. As defined in other studies \cite{Cardinot2016sab,Hauert2002},
abstainers receive a payoff greater than $P$ and less than $R$ (i.e., $P<L<R$).
Thus, considering the normalized payoff matrix adopted, $0<l<1$. The payoff
matrix and the associated values are illustrated in Table~\ref{tab:payoffs}.

\begin{table}[htb]
    \caption{The Optional Prisoner's Dilemma game matrix.}
    \label{tab:payoffs}
    \begin{subtable}{.4\linewidth}
        \centering
        \begin{tabular}{c c | c | c}
                & {\bf C} & {\bf D} & {\bf A} \\
            \cline{2-4}
            {\bf C}  & \multicolumn{1}{|c|}{\backslashbox{R}{R}}
                     & \backslashbox{S}{T}
                     & \multicolumn{1}{c|}{\backslashbox{L}{L}} \\
            \cline{1-4}
            {\bf D}  & \multicolumn{1}{|c|}{\backslashbox{T}{S}}
                     & \backslashbox{P}{P}
                     & \multicolumn{1}{c|}{\backslashbox{L}{L}} \\
            \cline{1-4}
            {\bf A}  & \multicolumn{1}{|c|}{\backslashbox{L}{L}}
                     & \backslashbox{L}{L}
                     & \multicolumn{1}{c|}{\backslashbox{L}{L}} \\
            \cline{2-4}
        \end{tabular}
        \caption{Extended game matrix.}
    \end{subtable}
    \begin{subtable}{.6\linewidth}
        \centering
        \setlength{\tabcolsep}{8pt}
        \begin{tabular}{l|c}
            {\bf Payoff} & {\bf Value} \\
            \hline
            {Temptation to defect (T)}              & $]1,2[$     \\
            {Reward for mutual cooperation (R)}     & $1$         \\
            {Punishment for mutual defection (P)}   & $0$         \\
            {Sucker's payoff (S)}                   & $0$         \\
            {Loner's payoff (L)}                    & $]0,1[$     \\
        \end{tabular}
        \caption{Payoff values.}
    \end{subtable}
\end{table}

In these experiments, the following parameters are used: a $102 \times 102$
~($N=102^2$) regular lattice grid with periodic boundary conditions is created
and fully populated with agents, which can play with their eight immediate
neighbours (Moore neighbourhood). We adopt an unbiased environment (in which
initially each agent is designated as a cooperator ($C$), defector ($D$) or
abstainer ($A$) with equal probability) as well as investigating a biased
environment (in which the percentage of abstainers present in the environment
is varied). Also, each edge linking agents has the same initial weight $w=1$,
which will adaptively change in accordance with the interaction.

Monte Carlo methods are used to perform the Coevolutionary Optional Prisoner's
Dilemma game. In one Monte Carlo (MC) step, each player is selected once on
average. This means that one MC step comprises $N$ inner steps where the
following calculations and updates occur:

\begin{itemize}
    \item Select an agent ($x$) at random from the population.

    \item Calculate the utility $u_{xy}$ of each interaction of $x$ with its
        eight neighbours (each neighbour represented as agent $y$) as follows:
        \begin{equation}
            u_{xy} = w_{xy} P_{xy},
        \end{equation}
        where $w_{xy}$ is the edge weight between agents $x$ and $y$, and
        $P_{xy}$ corresponds to the payoff obtained by agent $x$ on playing the
        game with agent $y$.

    \item Calculate $U_x$ the accumulated utility of $x$, that is:
        \begin{equation}
            U_x = \sum_{y \in \Omega_x}u_{xy},
        \end{equation}
        where $\Omega_x$ denotes the set of neighbours of the agent $x$.

    \item In order to update the link weights, $w_{xy}$, between agents, compare
        the values of $u_{xy}$ and the average accumulated utility
        (i.e., $\bar{U_x}=U_x/8$) as follows:
        \begin{equation}
            \label{eq:bigdelta}
            w_{xy} =
            \begin{dcases*}
                w_{xy} + \Delta  & if $u_{xy} > \bar{U_x}$ \\
                w_{xy} - \Delta  & if $u_{xy} < \bar{U_x}$ \\
                w_{xy}           & otherwise
            \end{dcases*},
        \end{equation}
        where $\Delta$ is a constant such that $0 \le \Delta \le \delta$.

    \item In line with previous research \cite{Huang2015,Wang2014}, $w_{xy}$
        is adjusted to be within the range
        \begin{equation}
            \label{eq:smalldelta}
            1-\delta \le w_{xy} \le 1+\delta,
        \end{equation}
        where $\delta$ ($0 < \delta \le 1$) defines the weight heterogeneity.
        Note that when $\Delta$ or $\delta$ are equal to $0$, the link weight
        keeps constant (${w=1}$), which results in the traditional scenario
        where only the strategies evolve.

    \item In order to update the strategy of $x$, the accumulated utility $U_x$
        is recalculated (based on the new link weights) and compared with the
        accumulated utility of one randomly selected neighbour ($U_y$).
        If $U_y>U_x$, agent $x$ will copy the strategy of agent $y$ with a
        probability proportional to the utility difference
        (Equation~\ref{eq:prob}), otherwise, agent $x$ will keep its strategy
        for the next step.
        \begin{equation}
            \label{eq:prob}
            p(s_x=s_y) = \frac{U_y-U_x}{8(T-P)},
        \end{equation}
        where $T$ is the temptation to defect and $P$ is the punishment for
        mutual defection. This equation has been considered previously by
        Huang et al. \cite{Huang2015}.
\end{itemize}

Simulations are run for $10^5$ MC steps and the fraction of cooperation is
determined by calculating the average of the final $10^3$ MC steps. To alleviate
the effect of randomness in the approach, the final results are obtained by
averaging $10$ independent runs.
The following scenarios are investigated:
\begin{itemize}

    \item The benefits of coevolution and abstention.

    \item Presence of abstainers in the coevolutionary model.

    \item Inspecting the coevolutionary environment.

    \item Investigating the properties of the parameters $\Delta$ and $\delta$.

    \item Varying the number of states.

    \item Investigating the relationship between $\Delta/\delta$, $b$ and $l$.

    \item Investigating the robustness of cooperation in a biased environment.

\end{itemize}

\section{The Benefits of Coevolution and Abstention}
\label{sec:results1}

This section presents some of the main differences between the outcomes
obtained by the proposed Coevolutionary Optional Prisoner's Dilemma (COPD) game
and other models which do not adopt the concept of coevolution and/or
abstention. In the COPD game, we also investigate how a population in an
unbiased environment evolves over time.

\subsection{Presence of Abstainers in the Coevolutionary Model}

In order to provide a means to effectively explore the impact of our
coevolutionary model, i.e., the Coevolutionary Prisoner's Dilemma (COPD) game,
in the emergence of cooperation, we start by investigating the performance of
some of the existing models.  Namely, the Coevolutionary Prisoner's Dilemma
(CPD) game (i.e., same coevolutionary model as the COPD but without the concept of
abstention), the traditional Prisoner's Dilemma (PD) game, and the Optional
Prisoner's Dilemma game.

\begin{figure}[t]
    \captionsetup[subfigure]{labelformat=empty}
    \centering
    \begin{subfigure}{0.325\textwidth}
        \centering
        \epsfig{file=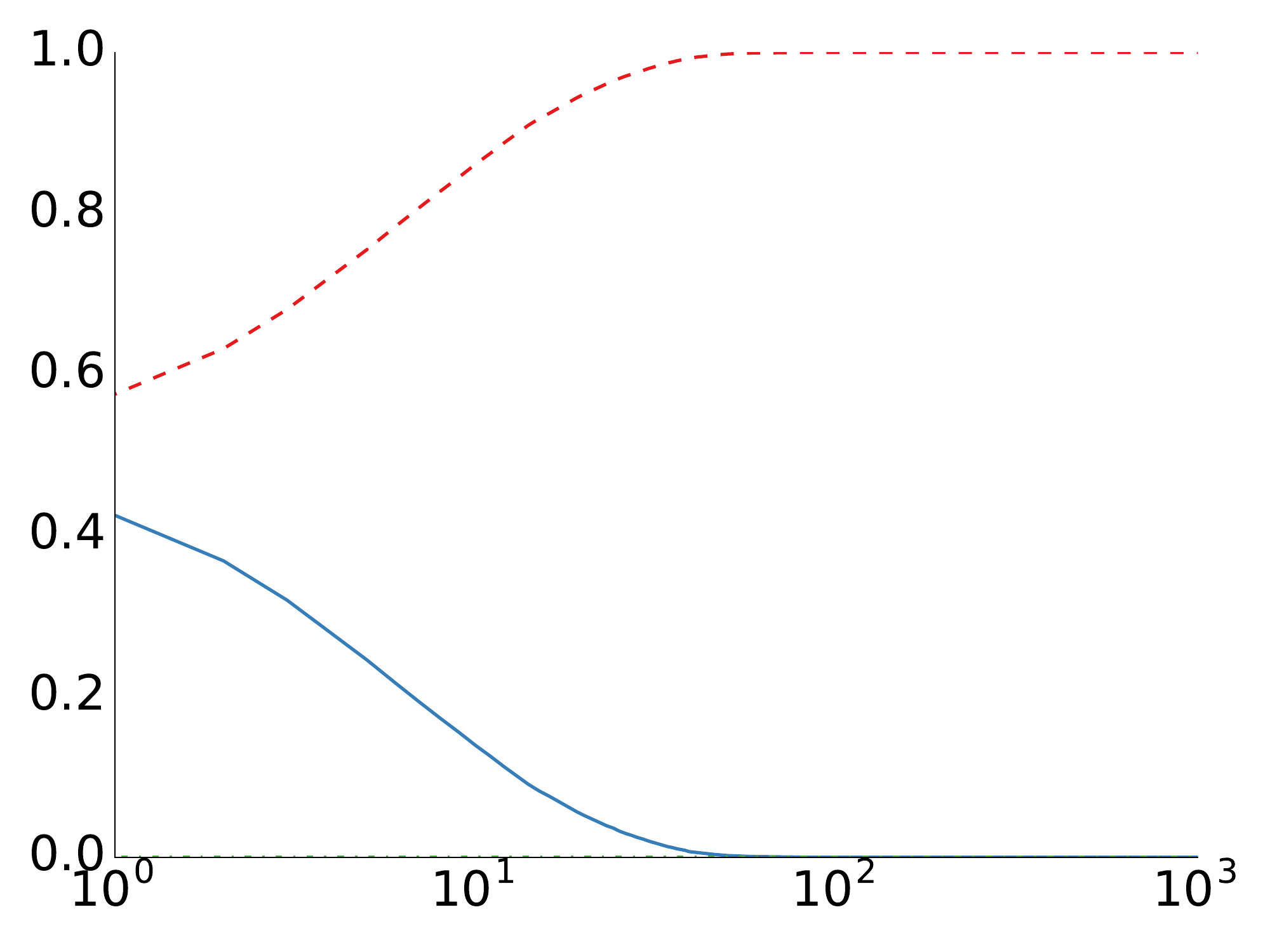, width=\textwidth}
        \caption{PD}
    \end{subfigure}
    \begin{subfigure}{0.325\textwidth}
        \centering
        \epsfig{file=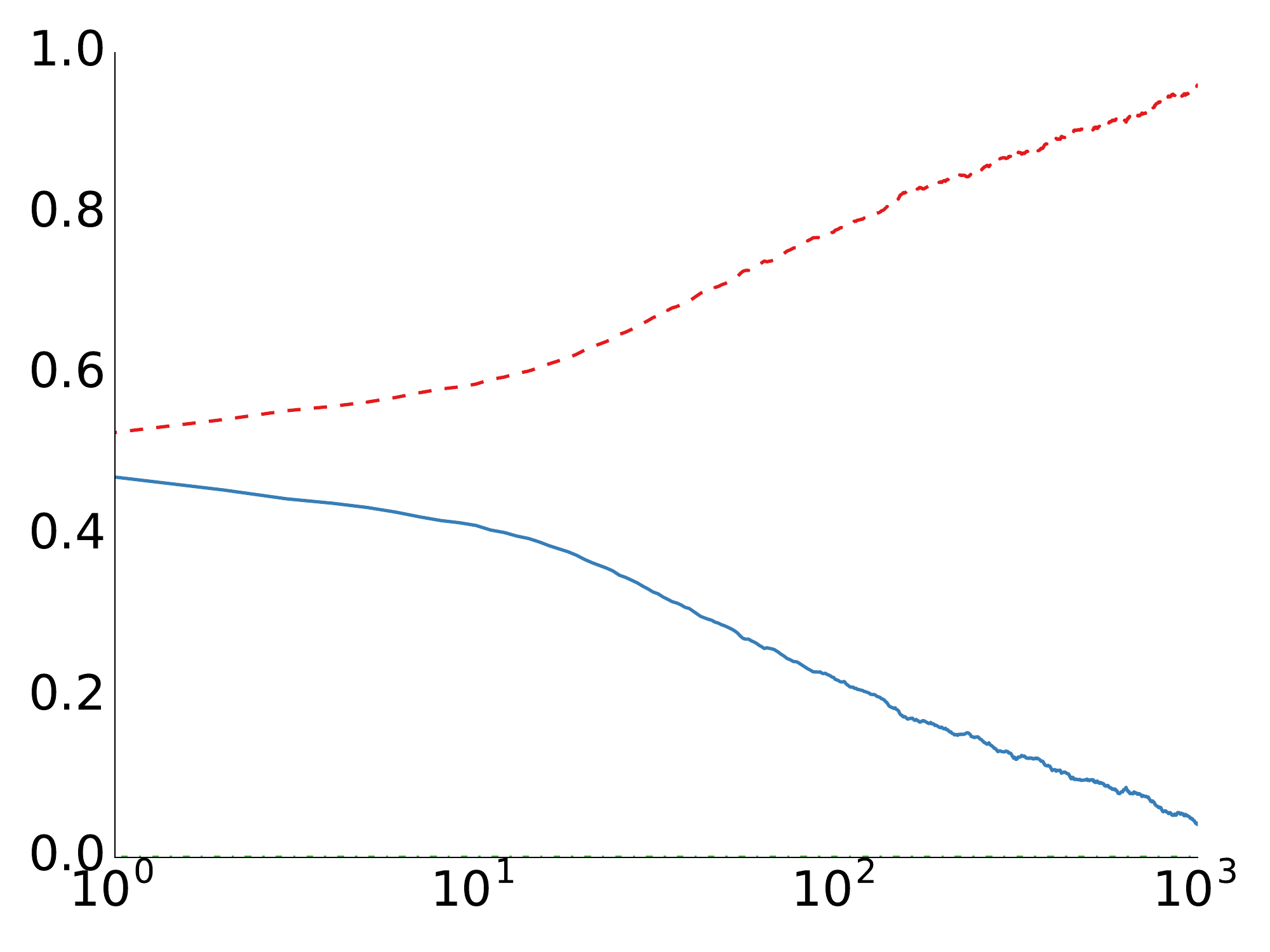, width=\textwidth}
        \caption{CPD (${\Delta=0.72;\ \delta=0.8}$)}
    \end{subfigure}
    \begin{subfigure}{0.325\textwidth}
        \centering
        \epsfig{file=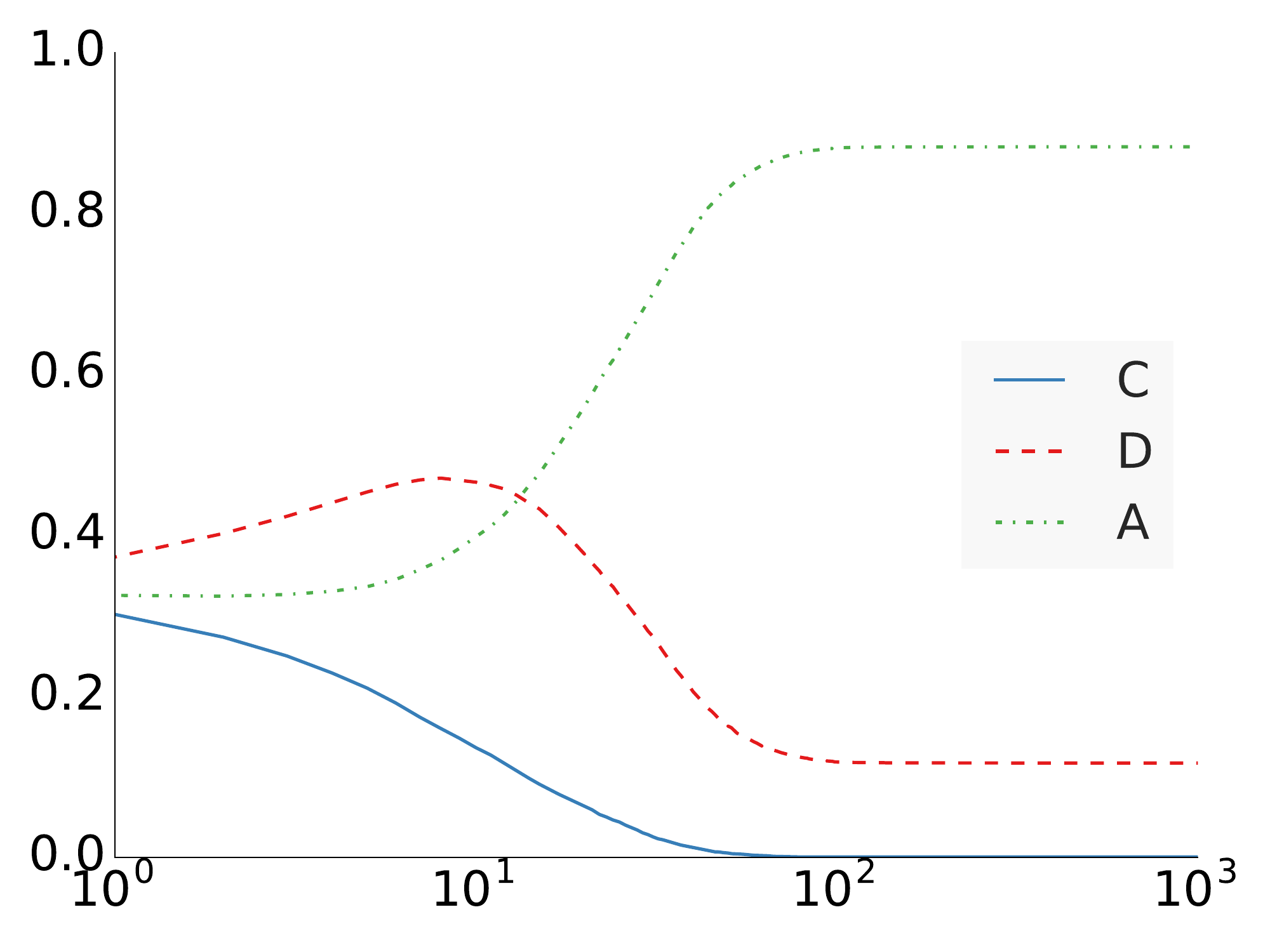, width=\textwidth}
        \caption{OPD ($l=0.6$)}
    \end{subfigure}
    \caption{
        Comparison of the Prisoner's Dilemma (PD), the Coevolutionary Prisoner's
        Dilemma (CPD) and the Optional Prisoner's Dilemma (OPD) games. All with the
        same temptation to defect, $b=1.9$.
    }
    \label{fig:compare}
\end{figure}

As shown in Figure~\ref{fig:compare}, it can be observed that for both PD and
CPD games, when the defector's payoff is very high (i.e., $b > 1.7$) defectors
spread quickly and dominate the environment. On the other hand, when abstainers
are present in a static and unweighted network, i.e., playing the OPD game, we
end up with abstainers dominating the environment. Undoubtedly, in many
scenarios, having a population of abstainers is better than a population of
defectors. However, it provides clear evidence that all these three models fail
to sustain cooperation.  In fact, results show that in this type of adverse
environment (i.e., with a high temptation to defect), cooperation has no
chance of emerging.

\begin{figure}[htb]
    \centering
    {\epsfig{file=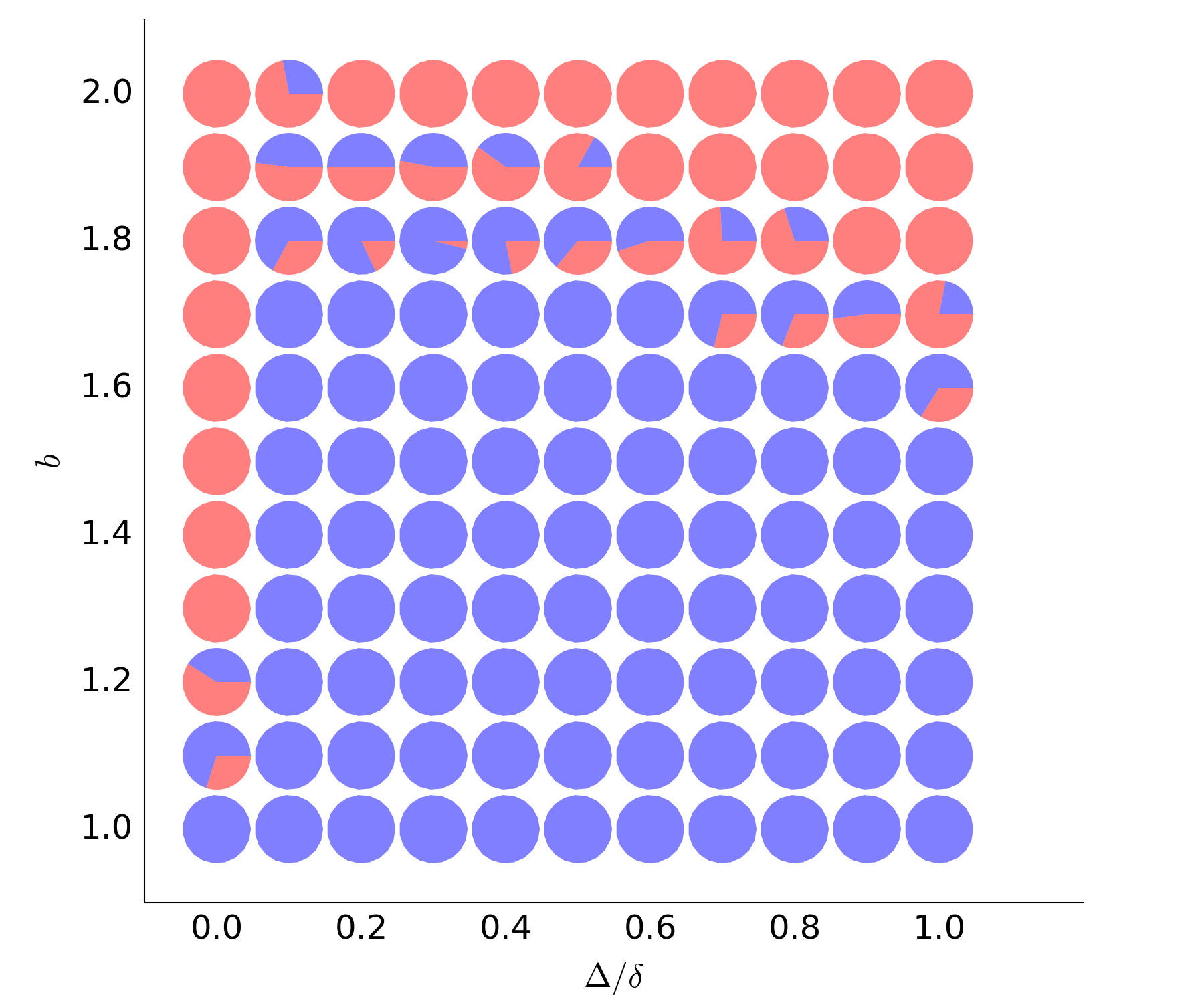, width=0.492\textwidth}}
    {\epsfig{file=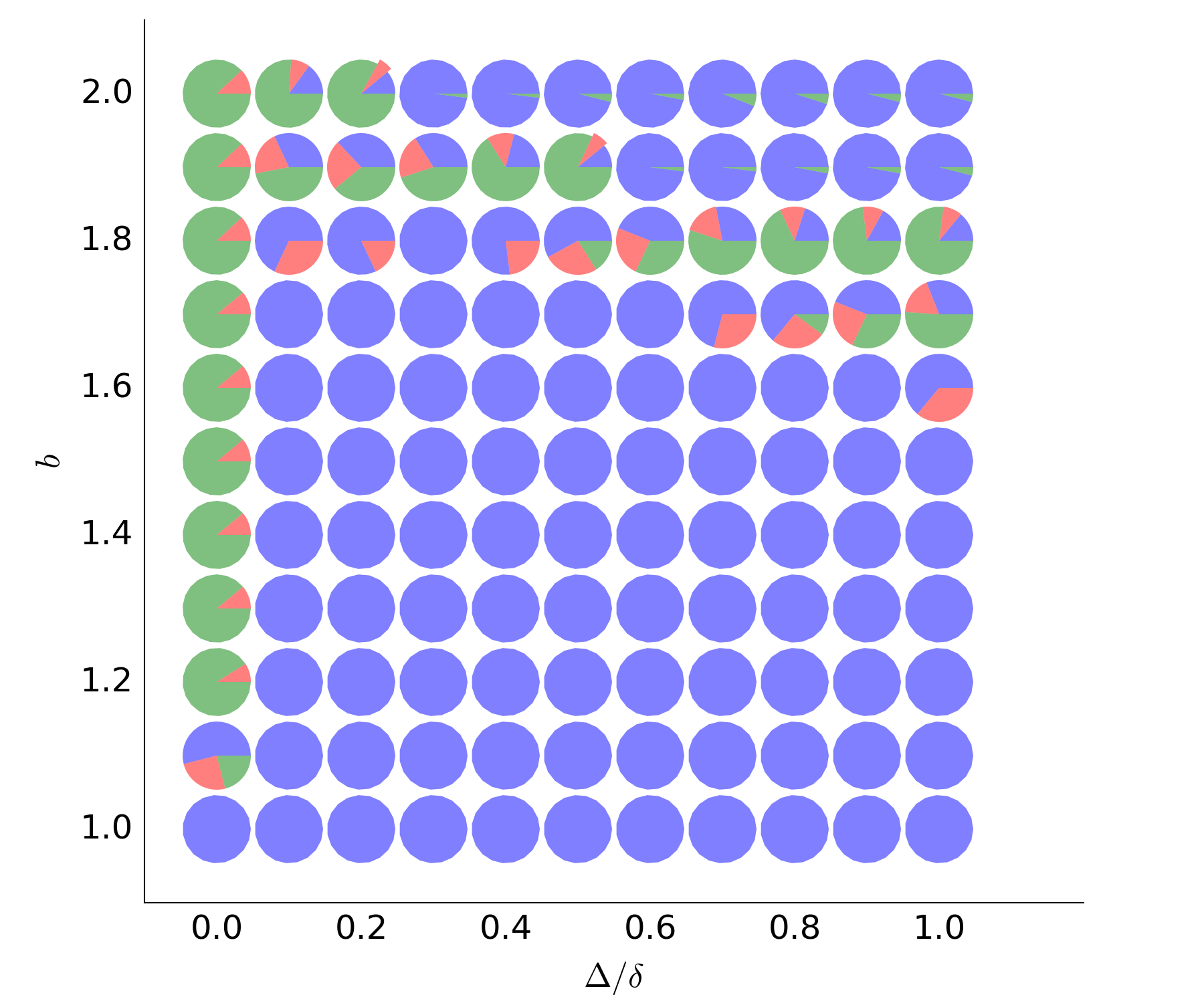, width=0.492\textwidth}}
    \caption{
        Typical phase diagram for an initial balanced population playing the
        Coevolutionary Prisoner's Dilemma game (left) and the Coevolutionary
        Optional Prisoner's Dilemma game with $l=0.6$ (right), both with $\delta=0.8$.
    }
    \label{fig:phase}
\end{figure}

Surprisingly, as shown in Figure~\ref{fig:phase}, when considering the
Coevolutionary Optional Prisoner's Dilemma (COPD) game for the same
environmental settings of Figure~\ref{fig:compare} (i.e., $l=0.6$,
$\Delta=0.72$ and $\Delta=0.72$), with the temptation of defection almost at
its peak (i.e., $b=1.9$), it is possible to reach high levels of cooperation.

Figure~\ref{fig:phase} shows a typical phase diagram for both CPD and COPD
games for a fixed value of $\delta=0.8$ and $l=0.6$ (on the COPD game). It can be
observed that if a given environmental setting (i.e, $b$, $\Delta$ and
$\delta$) produces a stable population of cooperators in the CPD game, then the
presence of abstainers will not change it. In other words, the COPD game does
not affect the outcome of scenarios in which cooperation is stable in the
absence of abstainers. Thus, the main changes occur in scenarios in which
defection dominates or survives ($b>1.5$).

To summarize, despite the fact that the Coevolutionary Prisoner's Dilemma (CPD)
game succeeds in the promotion of cooperation in a wide range of scenarios,
it is still not able to avoid the invasion by defectors in cases where $b>1.5$, which
does not happen when the abstainers are present (i.e., COPD game).

\subsection{Inspecting the Coevolutionary Environment}

\begin{figure}[p]
    \centering
    {\epsfig{file=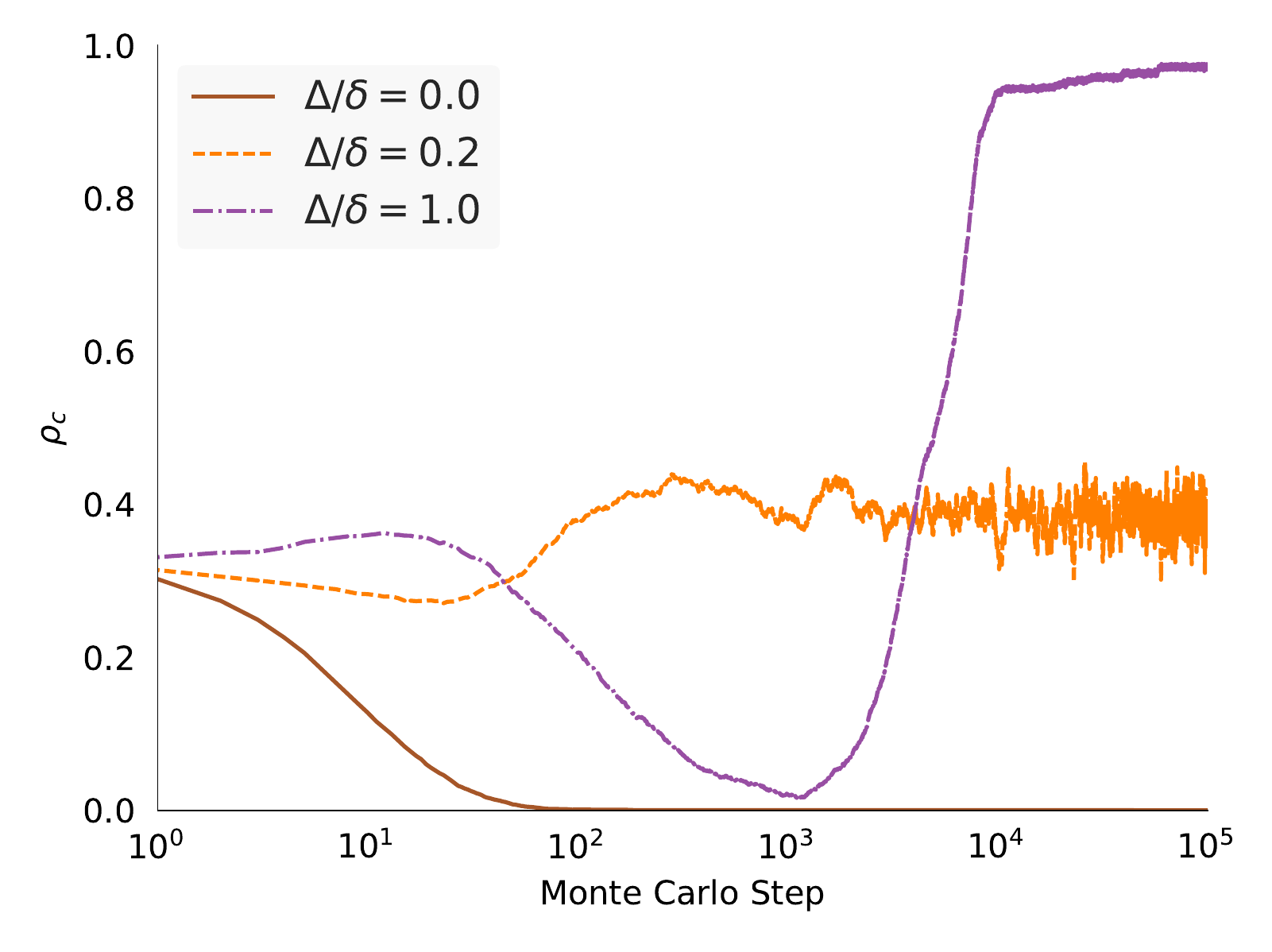, width=0.65\linewidth}}
    \caption{
        Progress of the fraction of cooperation $\rho_c$ during a Monte Carlo
        simulation for $b=1.9$, $l=0.6$ and $\delta=0.8$.
    }
    \label{fig:c_mcs}
\end{figure}

\begin{figure}[p]
    \centering
    {\epsfig{file=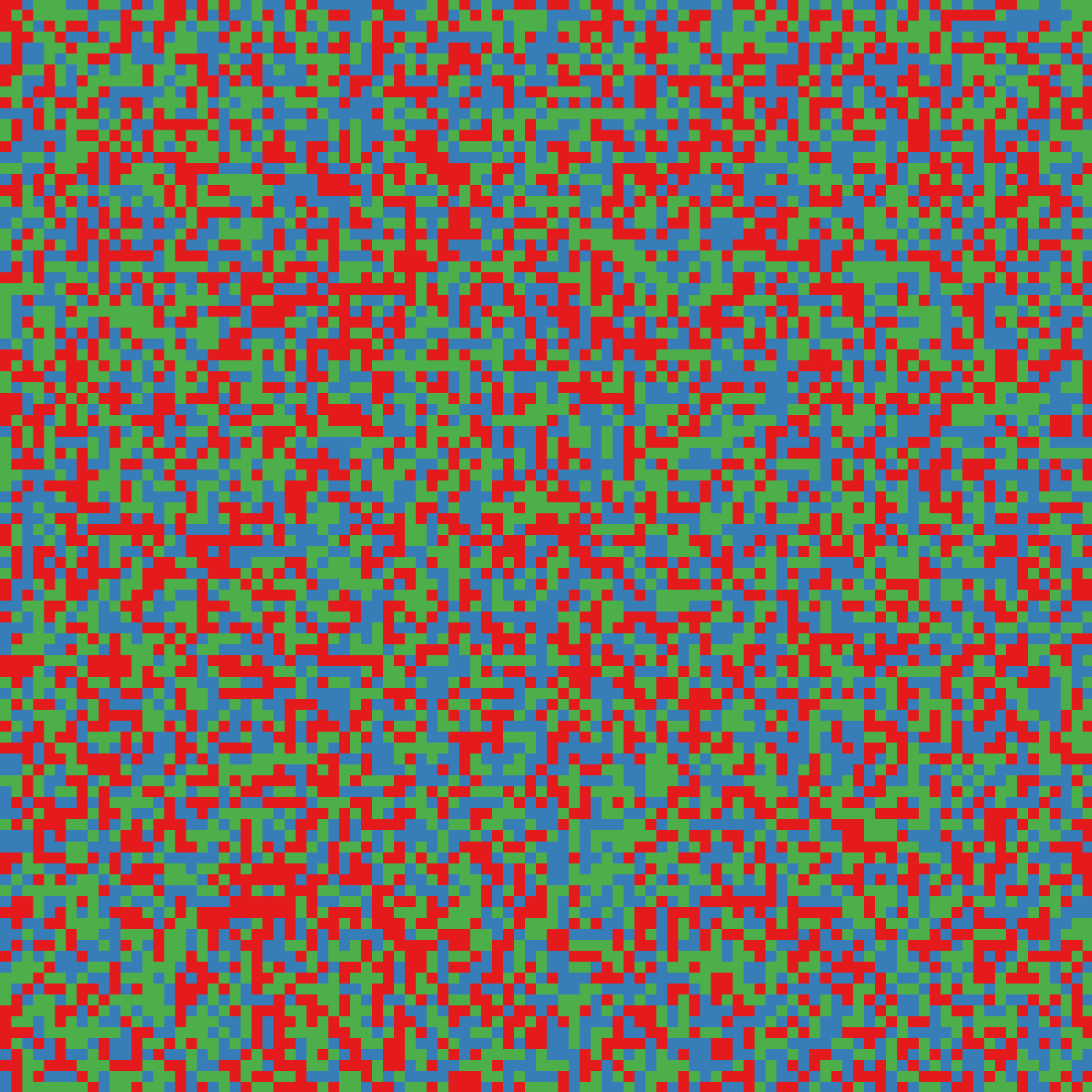, width=0.244\linewidth}}
    {\epsfig{file=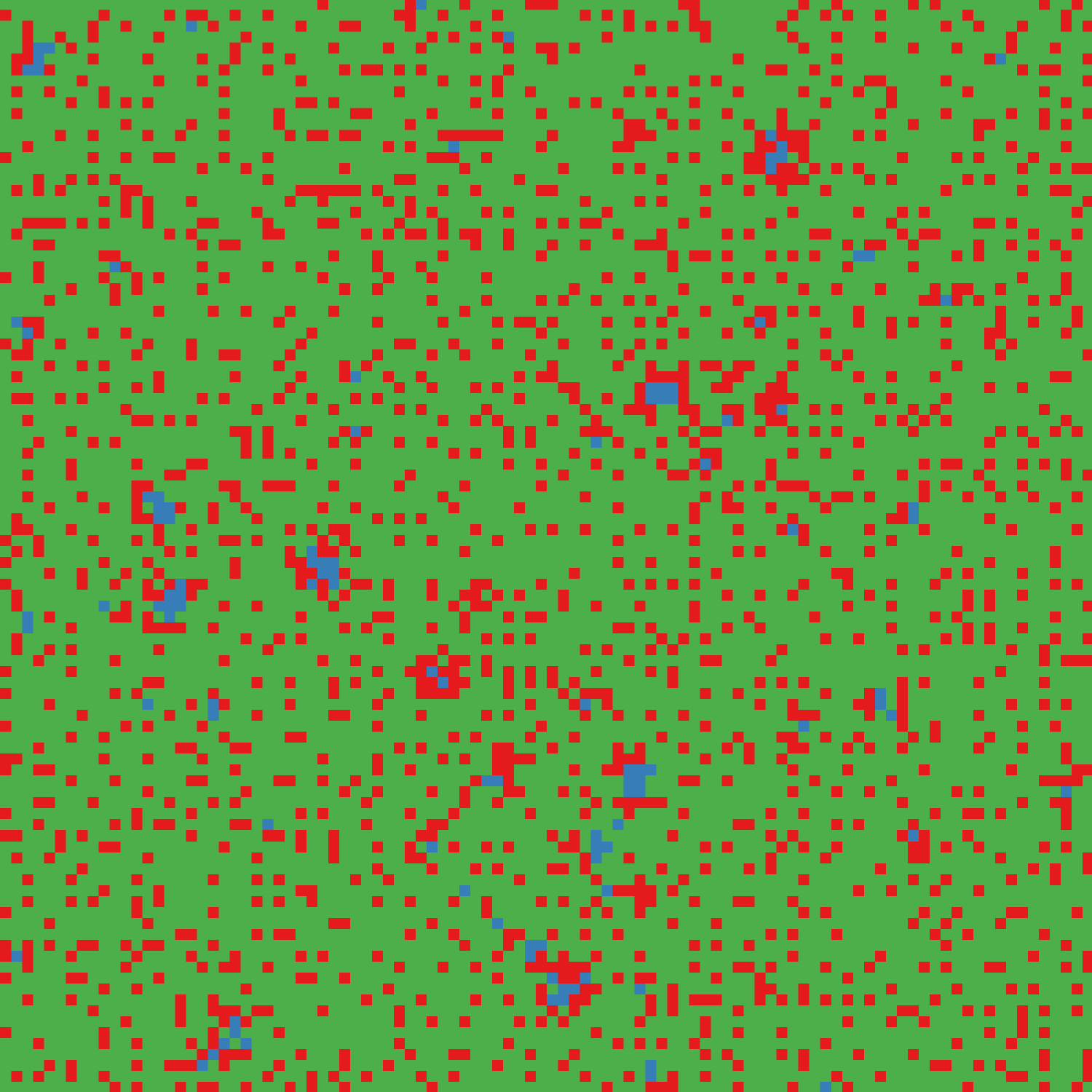, width=0.244\linewidth}}
    {\epsfig{file=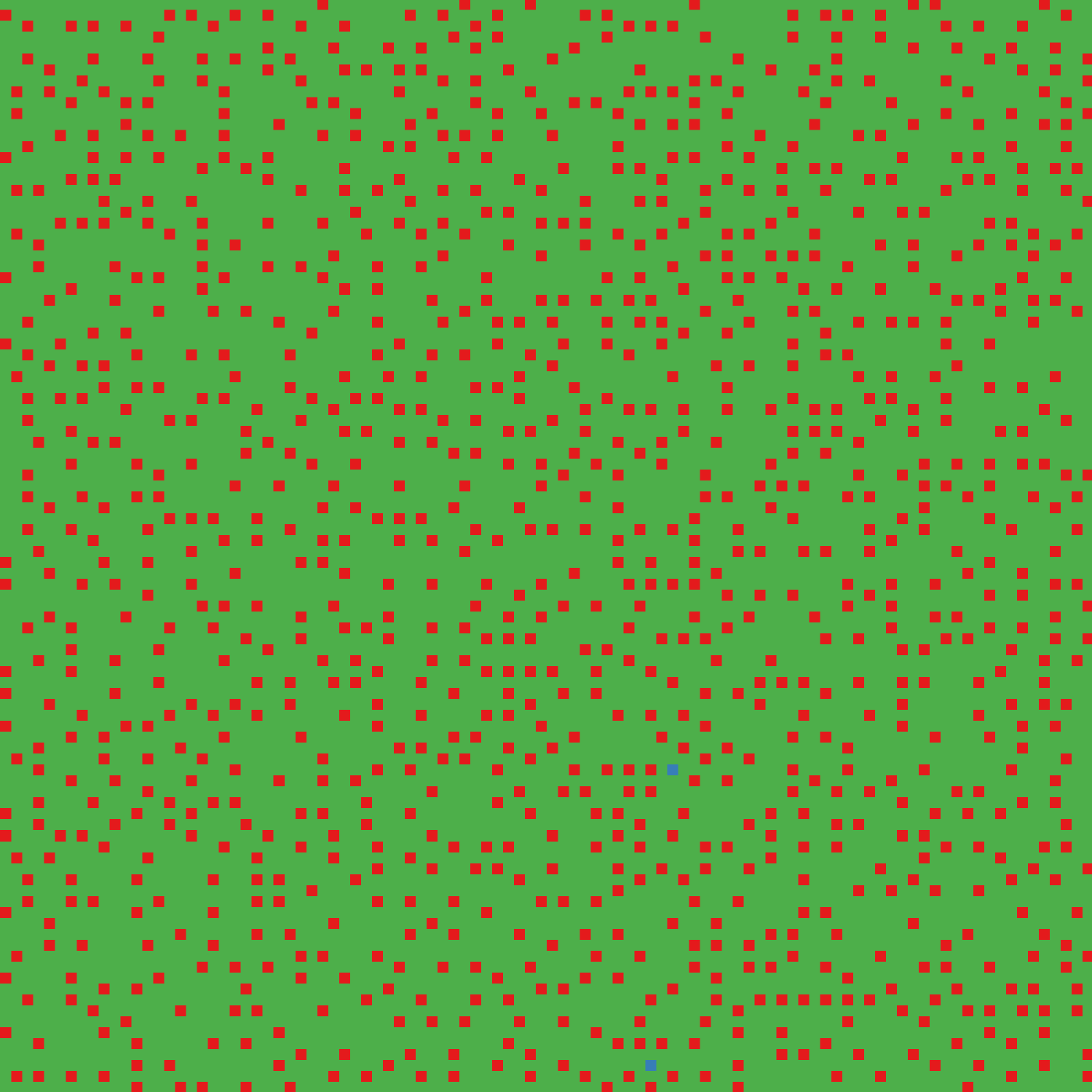, width=0.244\linewidth}}
    {\epsfig{file=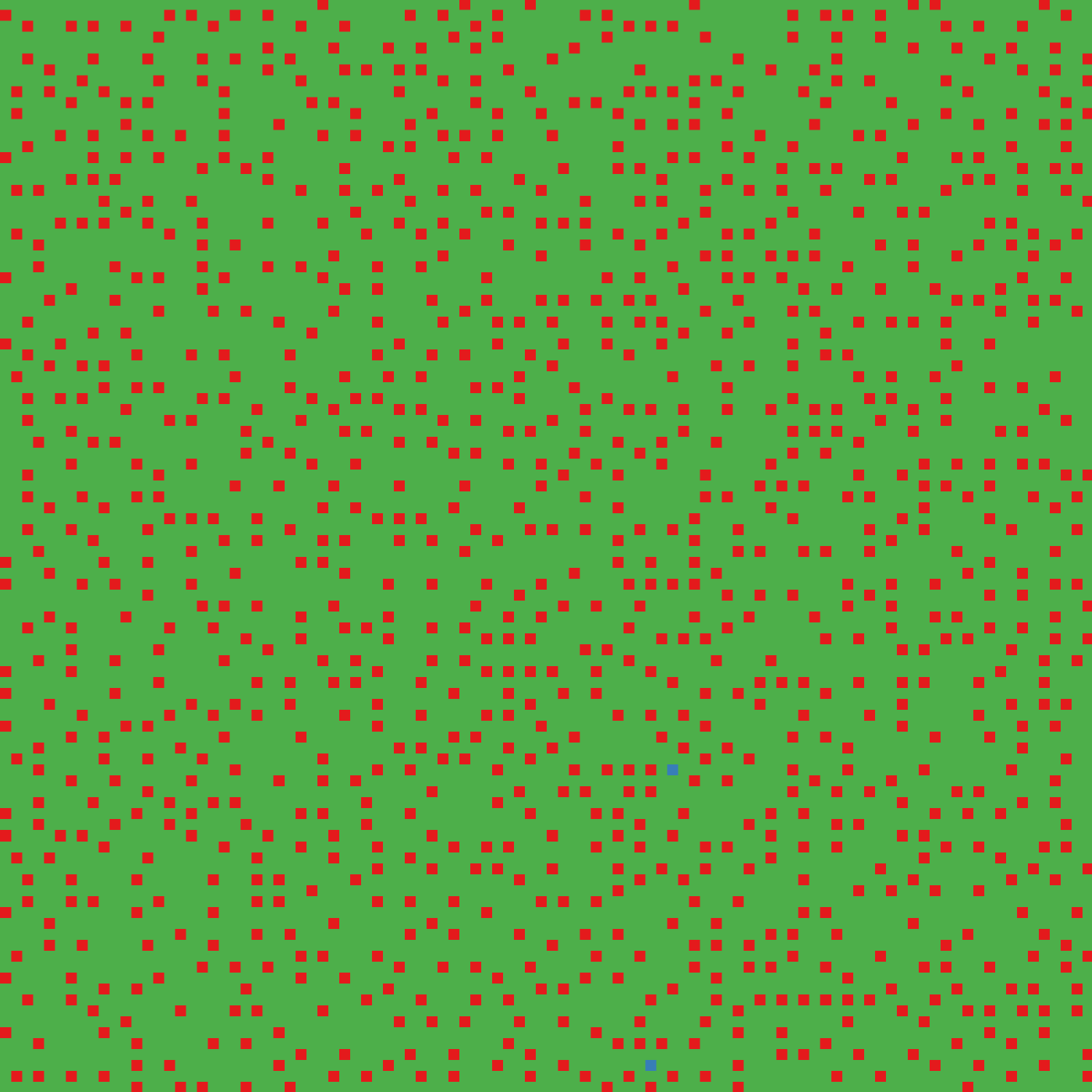, width=0.244\linewidth}}

    \vspace{0.07cm}

    {\epsfig{file=step0, width=0.244\linewidth}}
    {\epsfig{file=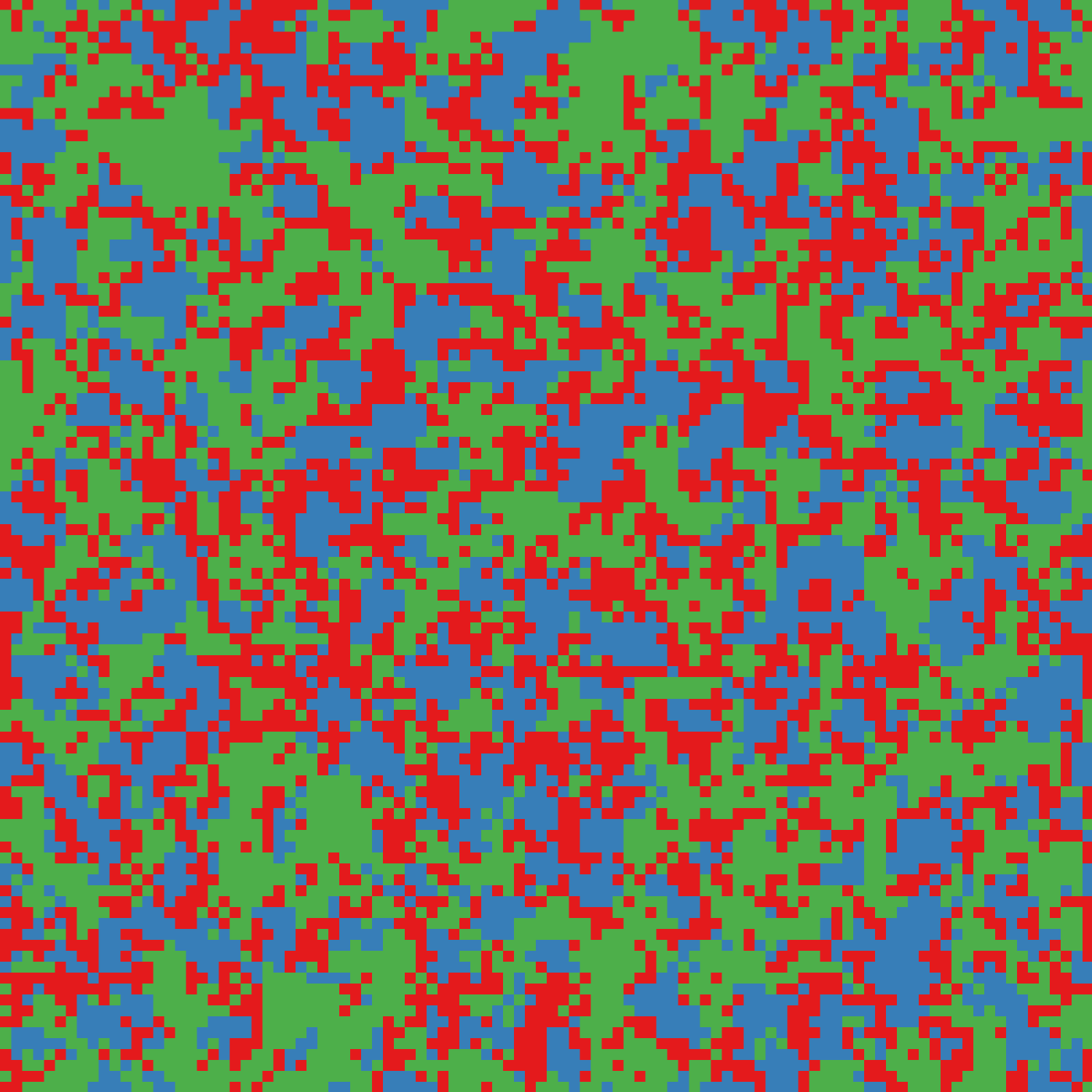, width=0.244\linewidth}}
    {\epsfig{file=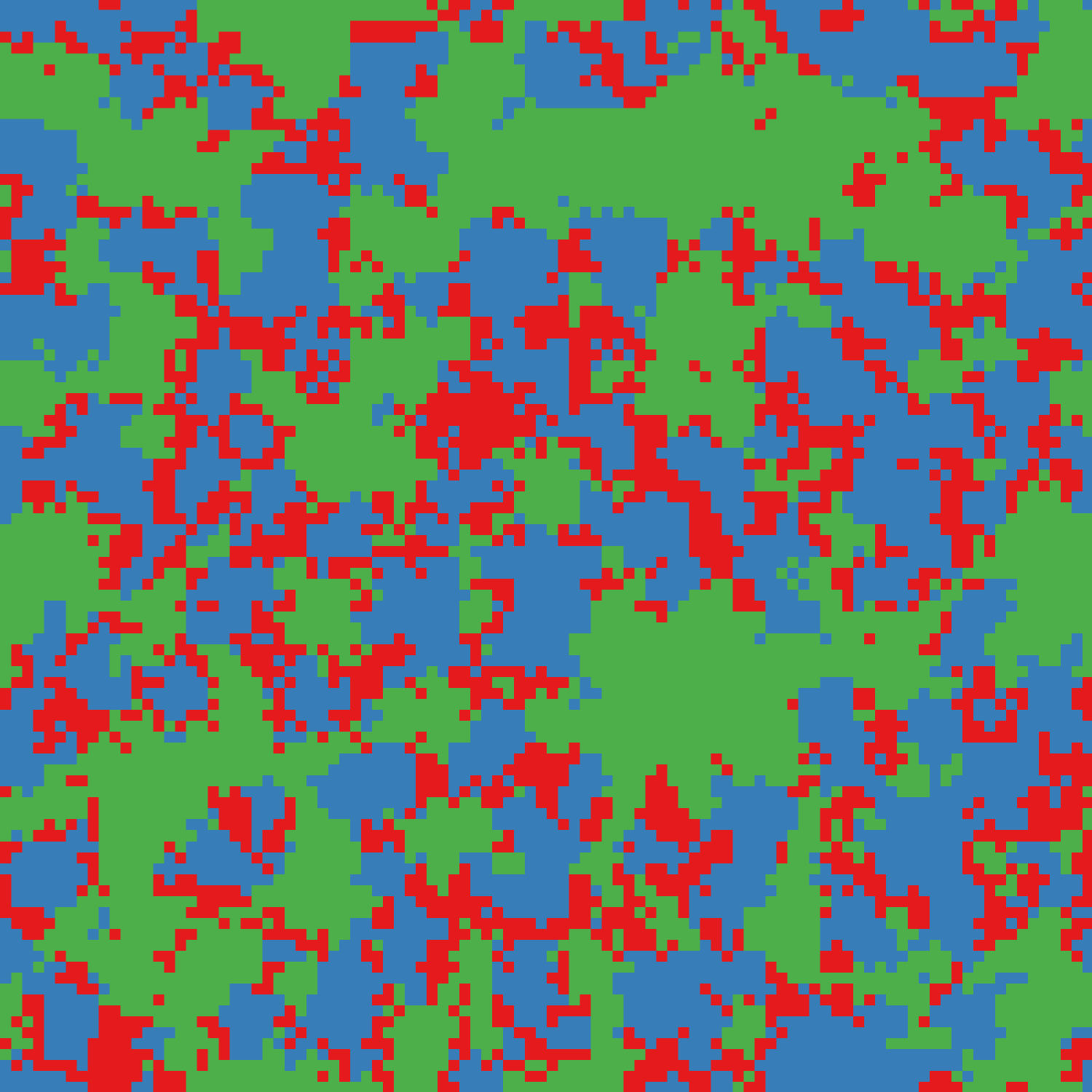, width=0.244\linewidth}}
    {\epsfig{file=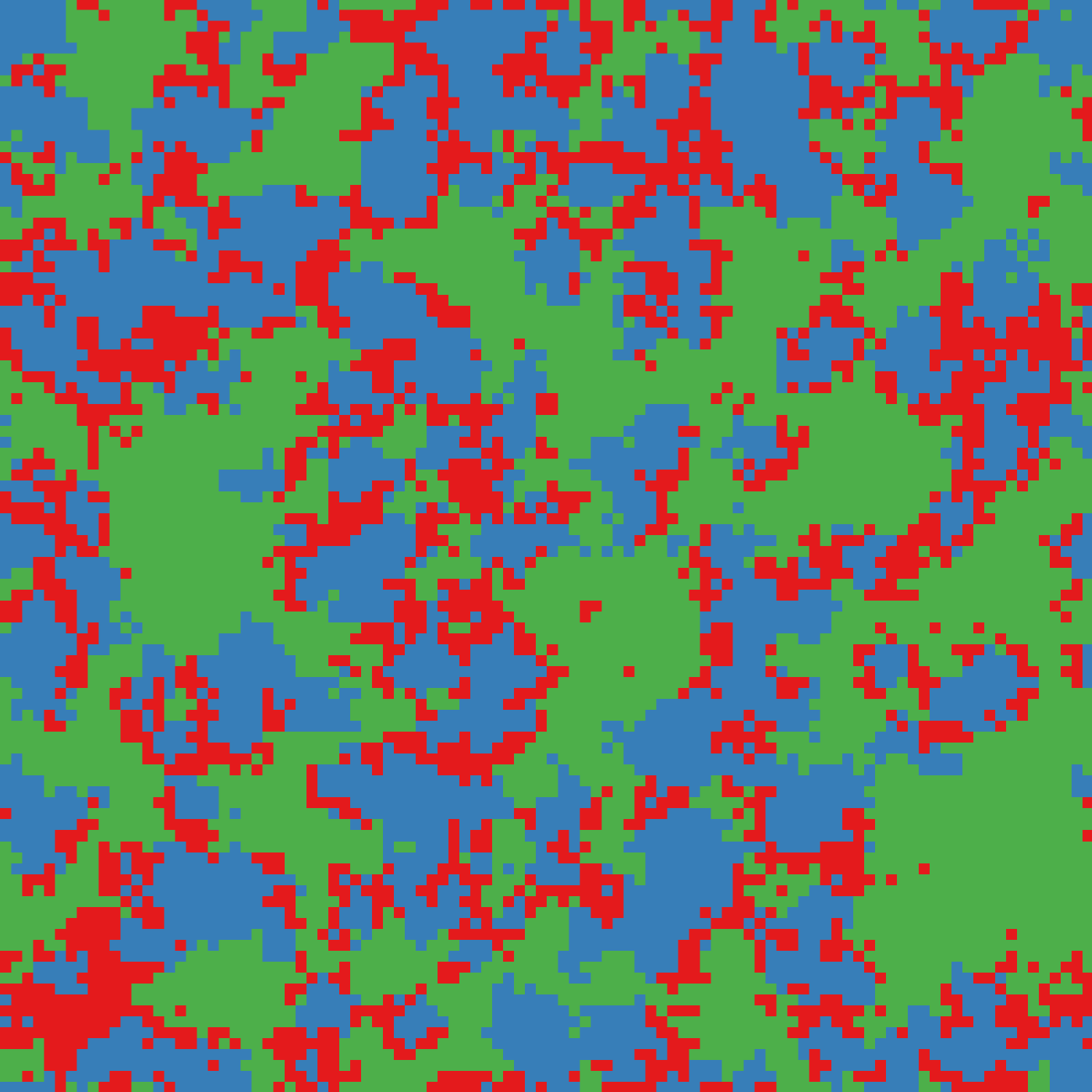, width=0.244\linewidth}}

    \vspace{0.07cm}

    {\epsfig{file=step0, width=0.244\linewidth}}
    {\epsfig{file=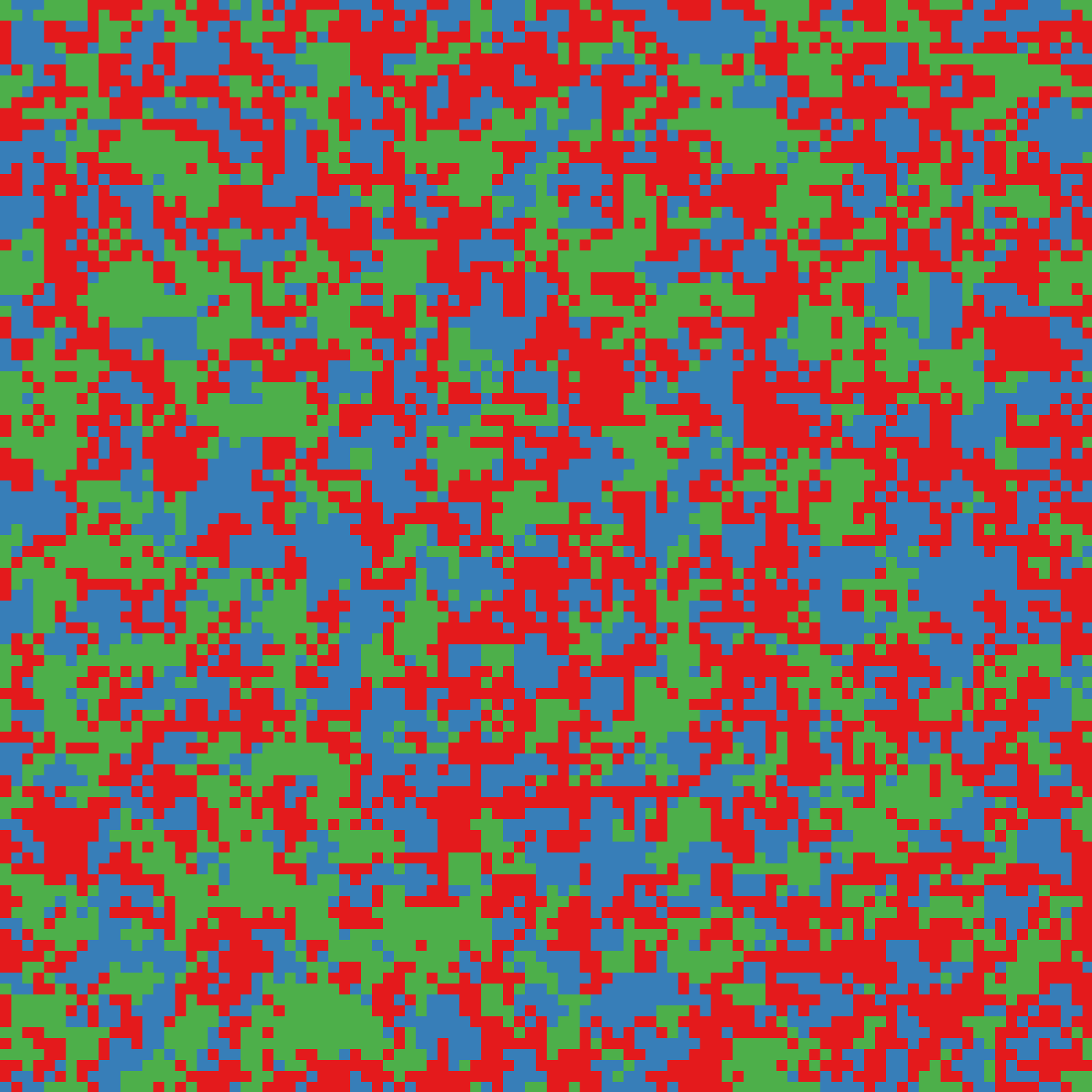, width=0.244\linewidth}}
    {\epsfig{file=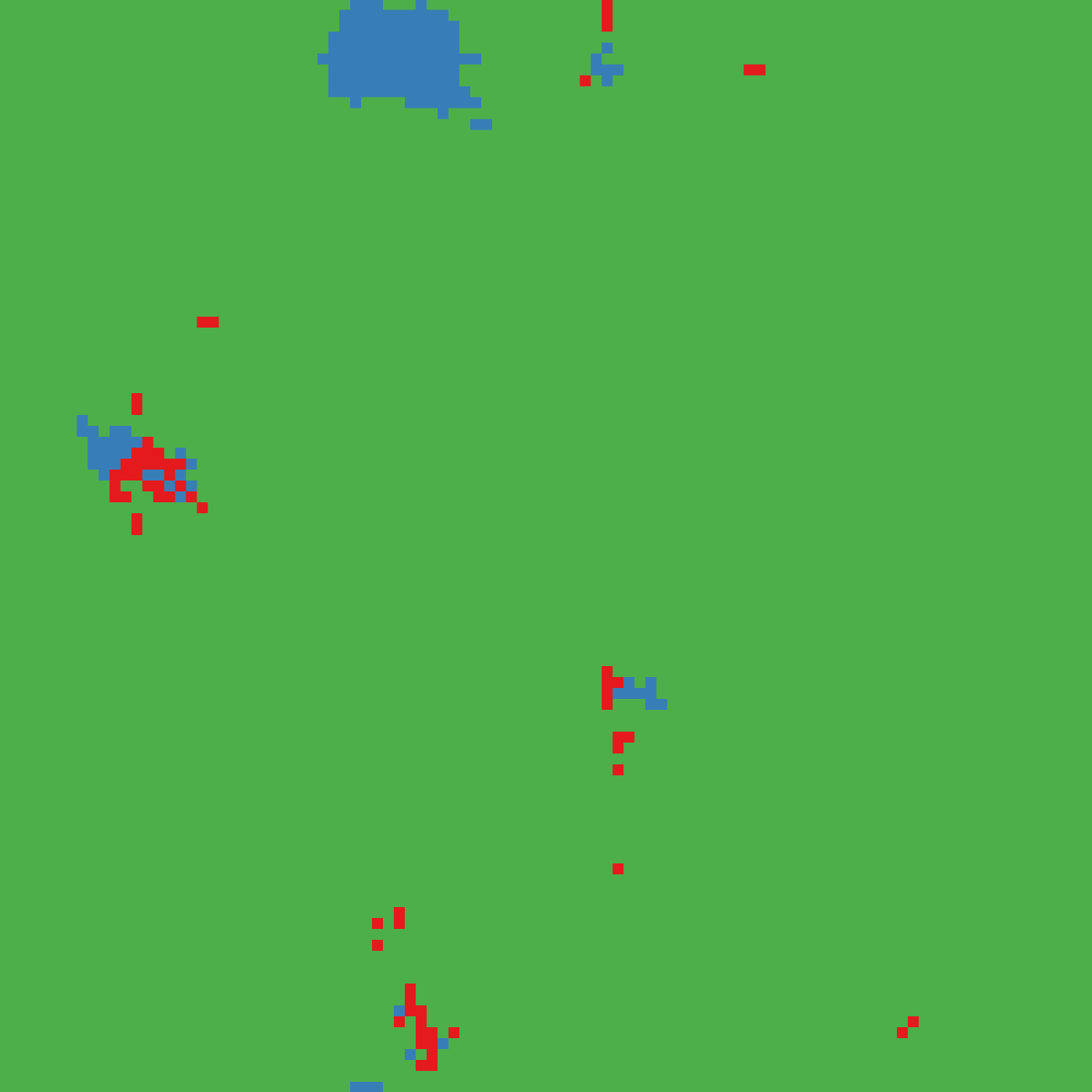, width=0.244\linewidth}}
    {\epsfig{file=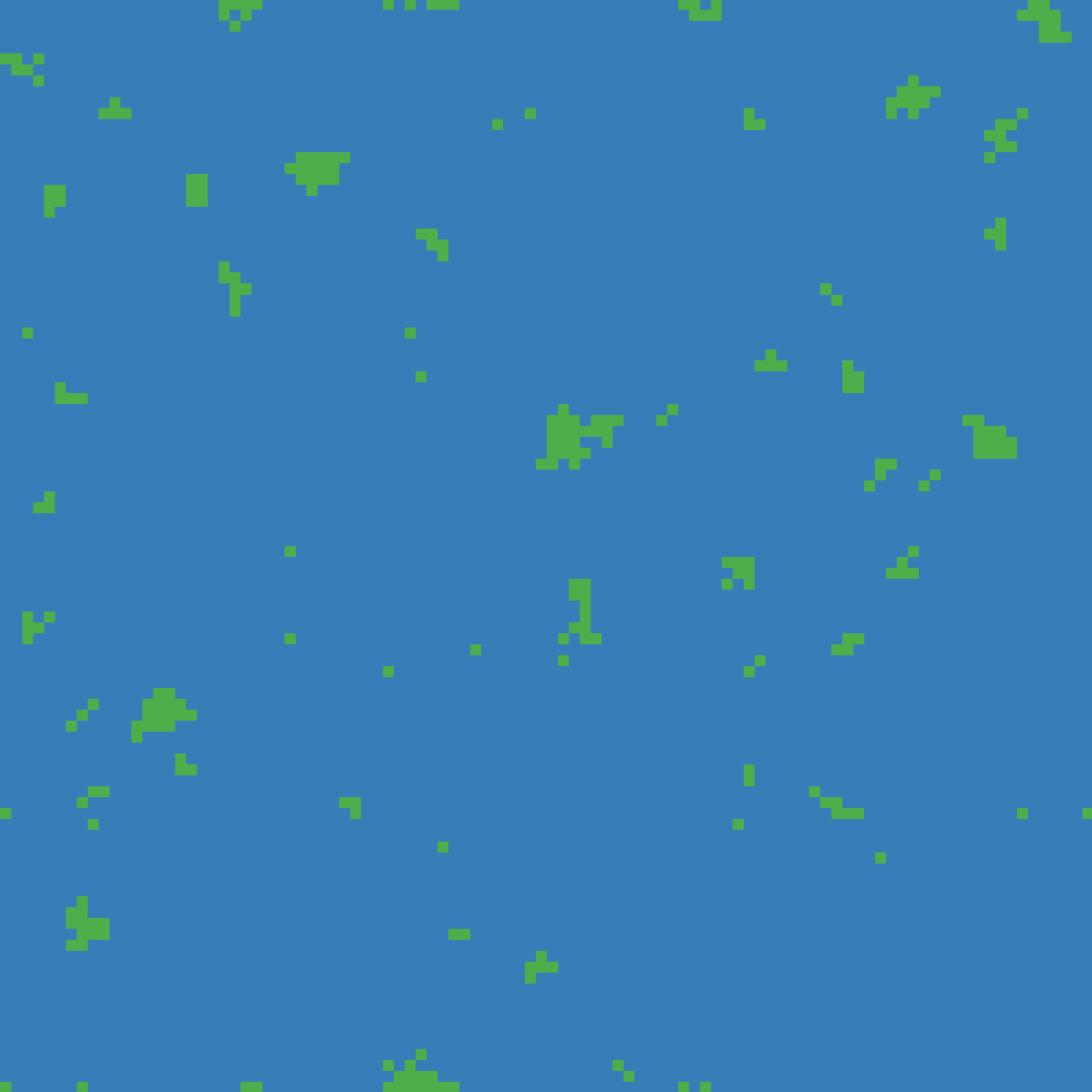, width=0.244\linewidth}}
    \caption{
        Snapshots of the distribution of the strategy in the Monte Carlo steps
        $0$, $45$, $1113$ and $10^5$ (from left to right) for $\Delta/\delta$
        equal to $0.0$, $0.2$ and $1.0$ (from top to bottom). In this Figure,
        cooperators, defectors and abstainers are represented by the colours
        blue, red and green respectively. All results are obtained for $b=1.9$,
        $l=0.6$ and $\delta=0.8$.
    }
    \label{fig:snapshots}
\end{figure}

In order to further explain the results witnessed in the previous experiments,
we investigate how the population evolves over time for the Coevolutionary
Optional Prisoner's Dilemma game.  Figure~\ref{fig:c_mcs} features the time
course of cooperation for three different values of $\Delta/\delta=\{0.0,\
0.2,\ 1.0\}$, which are some of the critical points when $b=1.9$, $l=0.6$ and
$\delta=0.8$. Based on these results, in Figure~\ref{fig:snapshots} we show
snapshots for the Monte Carlo steps $0$, $45$, $1113$ and $10^5$ for the three
scenarios shown in Figure~\ref{fig:c_mcs}.

We see from Figure~\ref{fig:snapshots} that for the traditional case (i.e.,
$\Delta/\delta=0.0$), abstainers spread quickly and reach a stable state in
which single defectors are completely isolated by abstainers. In this way, as
the payoffs obtained by a defector and an abstainer are the same, neither will
ever change their strategy. In fact, even if a single cooperator survives up to
this stage, for the same aforementioned reason, its strategy will not change
either. In fact, the same behaviour is noticed for any value of $b>1.2$ and
$\Delta/\delta=0$ (COPD in Figure~\ref{fig:phase}).

When $\Delta/\delta=0.2$, it is possible to observe some sort of equilibrium
between the three strategies. They reach a state of cyclic competition in which
abstainers invade defectors, defectors invade cooperators and cooperators
invade abstainers.

This behaviour, of balancing the three possible outcomes, is very common in
nature where species with different reproductive strategies remain in
equilibrium in the environment. For instance, the same scenario was observed as
being responsible for preserving biodiversity in the neighbourhoods of the
\textit{Escherichia coli}, which is a bacteria commonly found in the lower
intestine of warm-blooded organisms. According to Fisher \cite{Fisher2008},
studies were performed with three natural populations mixed together, in which
one population produces a natural antibiotic but is immune to its effects; a
second population is sensitive to the antibiotic but can grow faster than the
third population; and the third population is resistant to the antibiotic.

Because of this balance, they observed that each population ends up
establishing its own territory in the environment, as the first population
could kill off any other bacteria sensitive to the antibiotic, the second
population could use their faster growth rate to displace the bacteria which
are resistant to the antibiotic, and the third population could use their
immunity to displace the first population.

Another interesting behaviour is noticed for $\Delta/\delta=1.0$. In this
scenario, defectors are dominated by abstainers, allowing a few clusters of
cooperators to survive. As a result of the absence of defectors, cooperators
invade abstainers and dominate the environment.

\section{Exploring the Coevolutionary Optional Prisoner's Dilemma game}
\label{sec:results2}

In this section, we present some of the relevant experimental results of the
Monte Carlo simulations of the Coevolutionary Optional Prisoner's Dilemma game
in an unbiased environment. That is, a well-mixed initial population with a
balanced amount of cooperators, defectors and abstainers.

\subsection{Investigating the properties of $\Delta$ and $\delta$}
\label{sec:delta}

This section aims to investigate the properties of the presented model
(Sect.~\ref{sec:methodology}) in regard to the parameters $\Delta$ and
$\delta$. These parameters play a key role in the evolutionary dynamics of this
model because they define the number of possible link weights that an agent is
allowed to have (i.e., they define the number of states).

Despite the fact that the number of states is discrete, the act of counting
them is not straightforward. For instance, when counting the number of
states between $1-\delta$ and $1+\delta$ for $\Delta=0.2$ and $\delta=0.3$, we
could incorrectly state that there are four possible states for this scenario
(i.e., $\{0.7,\ 0.9,\ 1.1,\ 1.3\}$).  However, considering that the link weights
of all edges are initially set to $w=1$, and due to the other constraints
(Equations \ref{eq:bigdelta} and \ref{eq:smalldelta}), the number of states is
actually seven (i.e., $\{0.7,\ 0.8,\ 0.9,\ 1.0,\ 1.1,\ 1.2,\ 1.3\}$).

\begin{figure}[tb]
    \centering
    {\epsfig{file=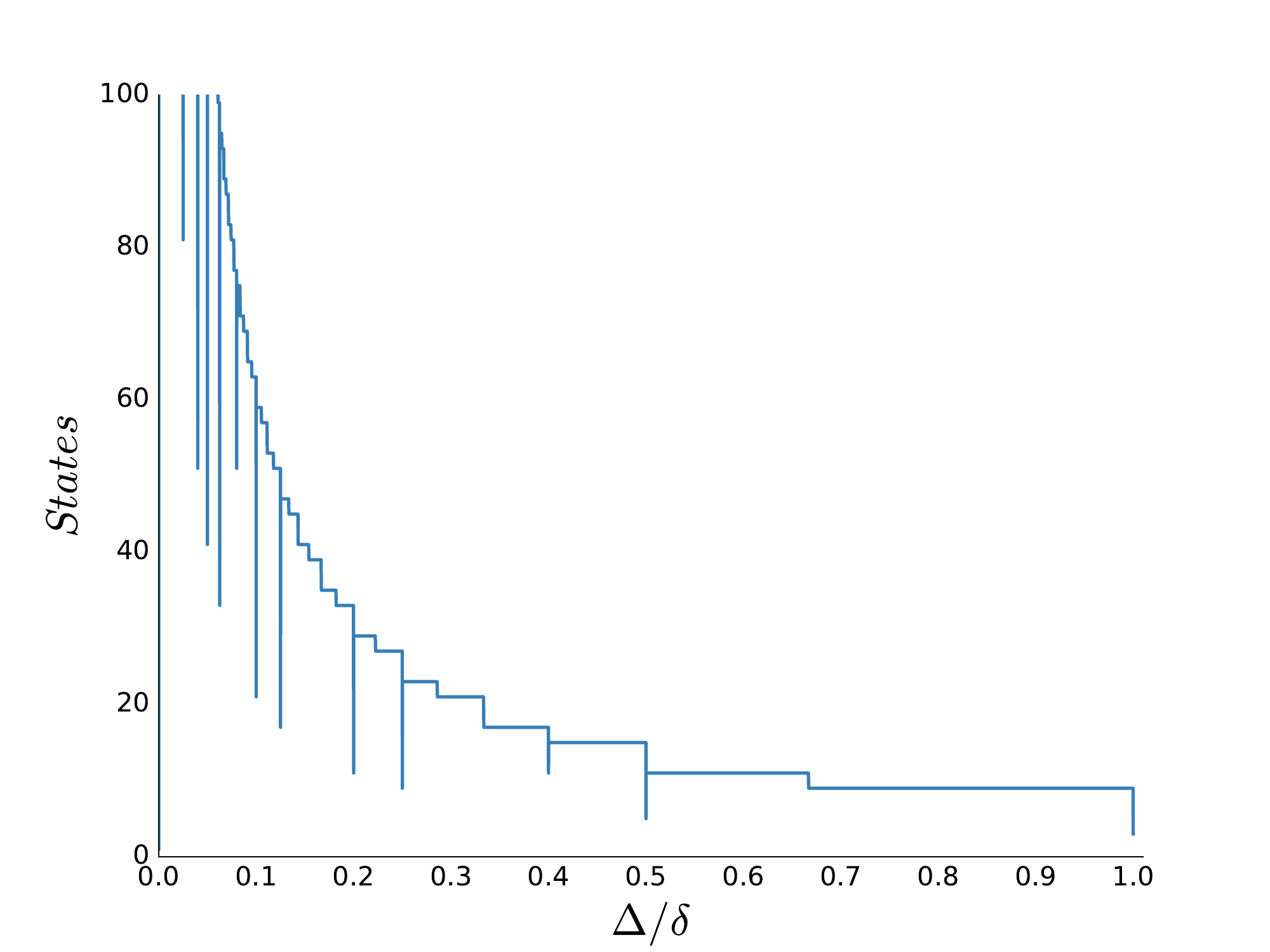, height=6.1cm}}
    \caption{
        The ratio $\Delta/\delta$ as a function of the number of states. For any combination
        of $\Delta$ and $\delta$, the ration $\Delta/\delta$ will always have the same number
        of states.
    }
    \label{fig:delta}
\end{figure}

In order to better understand the relationship between $\Delta$ and $\delta$,
we plot $\Delta$, $\delta$ and $\Delta/\delta$ as a function of the number of
states (numerically counted) for a number of different values of both
parameters (Figure~\ref{fig:delta}).  It was observed that given the pairs
$(\Delta_1,\ \delta_1)$ and $(\Delta_2/\delta_2)$, if $\Delta_1/\delta_1$ is
equal to $\Delta_2/\delta_2$, then the number of states of both settings is the
same.

Figure~\ref{fig:delta} shows the ratio $\Delta/\delta$ as a function of the
number of states.  As we can see, although the function is non-linear and
non-monotonic, in general, higher values of $\Delta/\delta$ have less states.

\subsection{Varying the Number of States}

\begin{figure}[tb]
    \centering
    {\epsfig{file=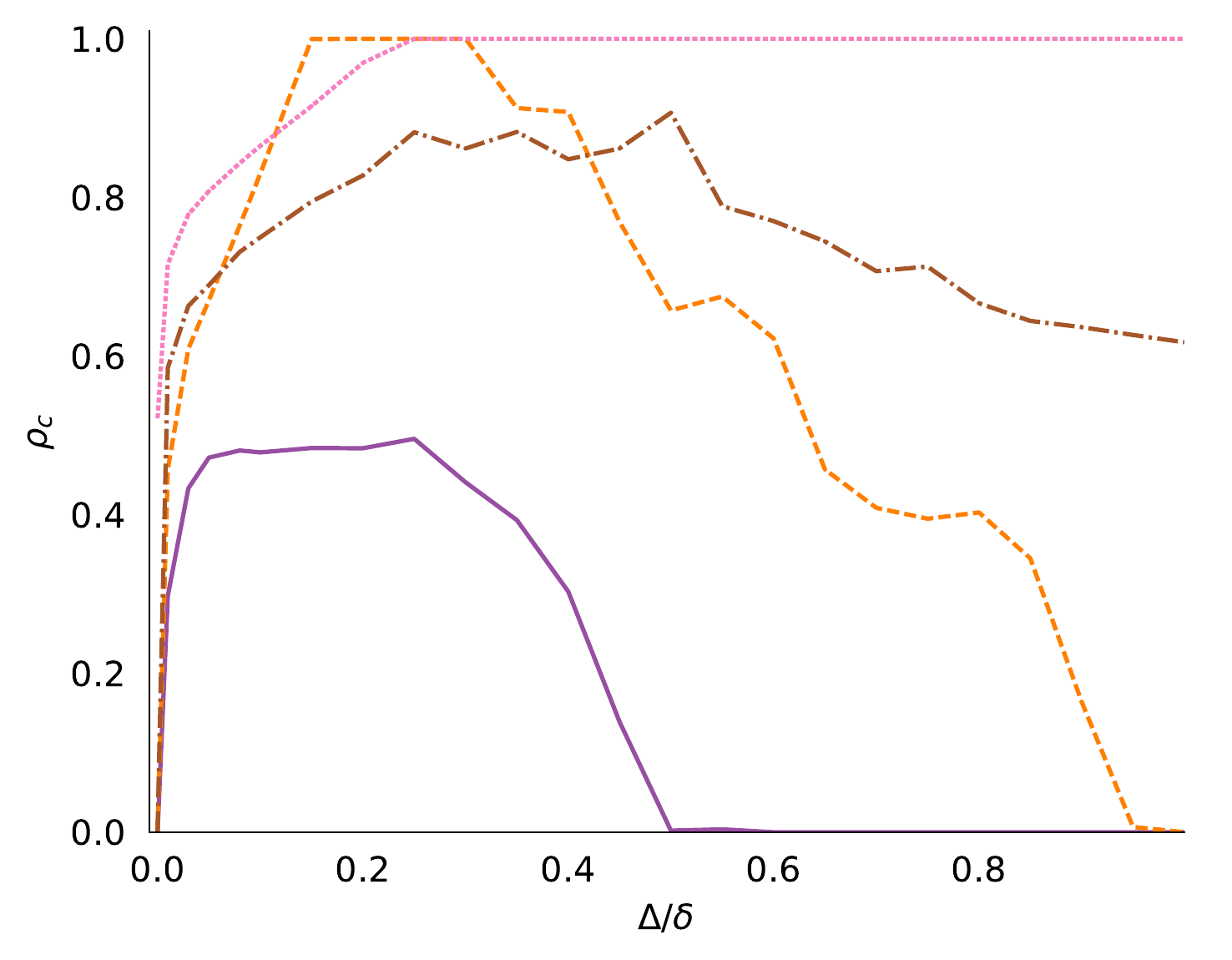, height=3.92cm}}
    {\epsfig{file=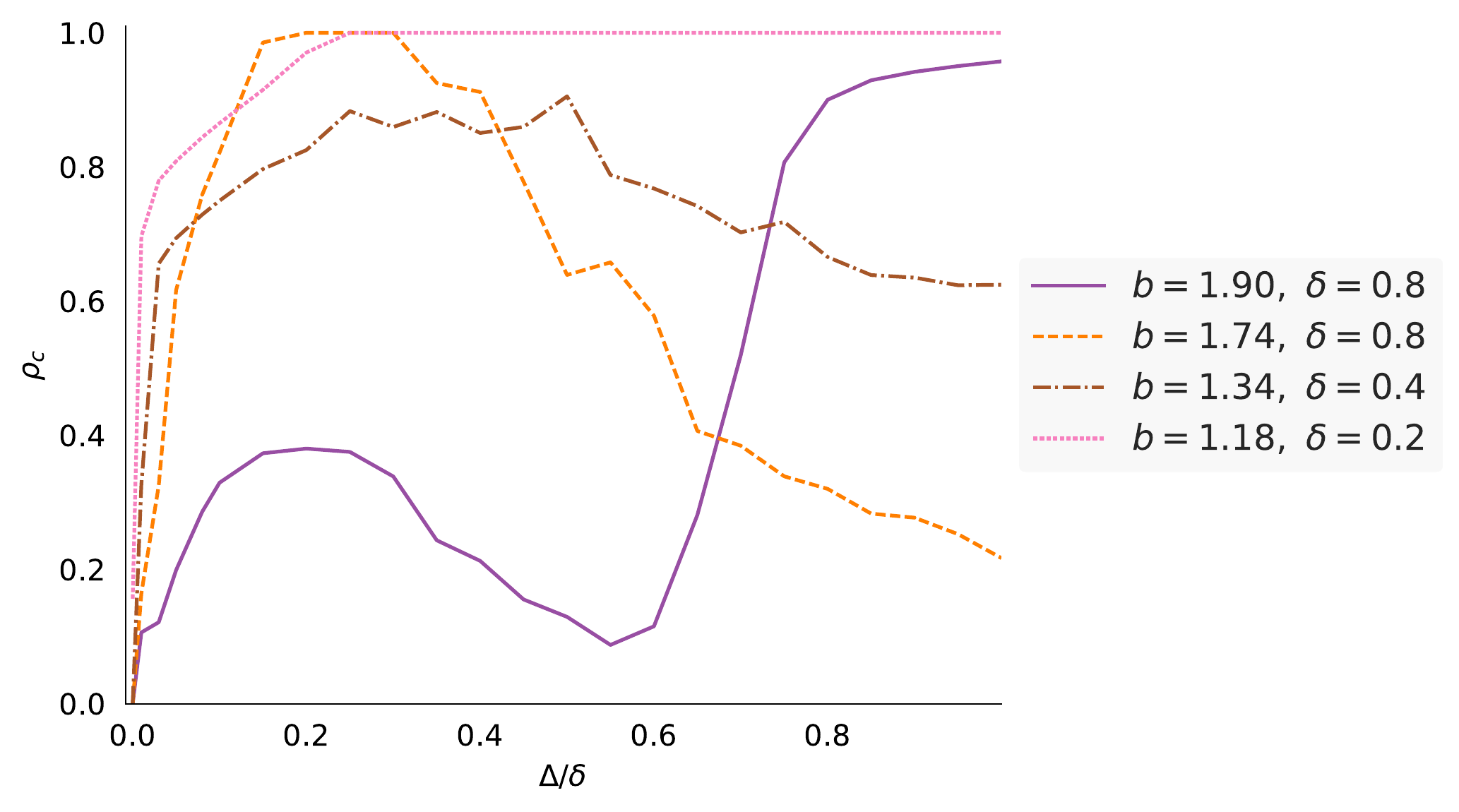, height=3.92cm}}
    \caption{
        Relationship between cooperation and the ratio $\Delta/\delta$ when
        the loner's payoff ($l$) is equal to $0.0$ (left) and $0.6$ (right).
    }
    \label{fig:c_amp}
\end{figure}

Figure~\ref{fig:c_amp} shows the impact of the coevolutionary model on the
emergence of cooperation when the ratio $\Delta/\delta$ varies for a range of
fixed values of the loner's payoff ($l$), temptation to defect ($b$) and
$\delta$. In this experiment, we observe that when $l=0.0$, the outcomes of the
Coevolutionary Optional Prisoner's Dilemma (COPD) game are very similar to
those observed by Huang et al. \cite{Huang2015} for the Coevolutionary
Prisoner's Dilemma (CPD) game. This result can be explained by the normalized
payoff matrix adopted in this work (Table~\ref{tab:payoffs}). Clearly,
when $l=0.0$, there is no advantage in abstaining from playing the game,
thus agents choose the option to cooperate or defect.

Results indicate that, in cases where the temptation to defect is very
low (e.g, $b \le 1.34$), the level of cooperation does not seem to be affected
by the increment of the loner's payoff, except when the advantage of abstaining
is very high (e.g, $l>0.8$). However, these results highlight that the presence
of the abstainers may protect cooperators from invasion. Moreover, the
difference between the traditional Optional Prisoner's Dilemma (i.e.,
$\Delta/\delta=0.0$) for $l=\{0.0,\ 0.6\}$ and all other values of
$\Delta/\delta$ is strong evidence that our coevolutionary model is very
advantageous to the promotion of cooperative behaviour.

Namely, when $l=0.6$, in the traditional case with a static and unweighted
network ($\Delta/\delta=0.0$), the cooperators have no chance of surviving;
except, of course, when $b$ is very close to the reward for mutual cooperation
$R$, where it is possible to observe scenarios of quasi-stable states of the
three strategies or between cooperators and defectors. In fact, in the
traditional OPD ($\Delta/\delta=0.0$), when $l>0.0$ and $b>1.2$, abstainers
are always the dominant strategy. However, when the coevolutionary rules are
used, cooperators do much better, being also able to dominate the whole
population in many cases.

It is noteworthy that the curves in Figure~\ref{fig:c_amp} are usually
non-linear and/or non-monotonic because of the properties of the ratio
$\Delta/\delta$ in regard to the number of states of each combination
of $\Delta$ and $\delta$ (Sect.~\ref{sec:delta}).

\subsection{Investigating the Relationship between $\Delta/\delta$, $b$ and $l$}
\label{sec:varyall}

To investigate the outcomes in other scenarios, we explore a wider range of
settings by varying the values of the temptation to defect ($b$), the loner's
payoff ($l$) and the ratio $\Delta/\delta$ for a fixed value of $\delta = 0.8$.

\begin{figure}[tb]
    \centering
    {\epsfig{file=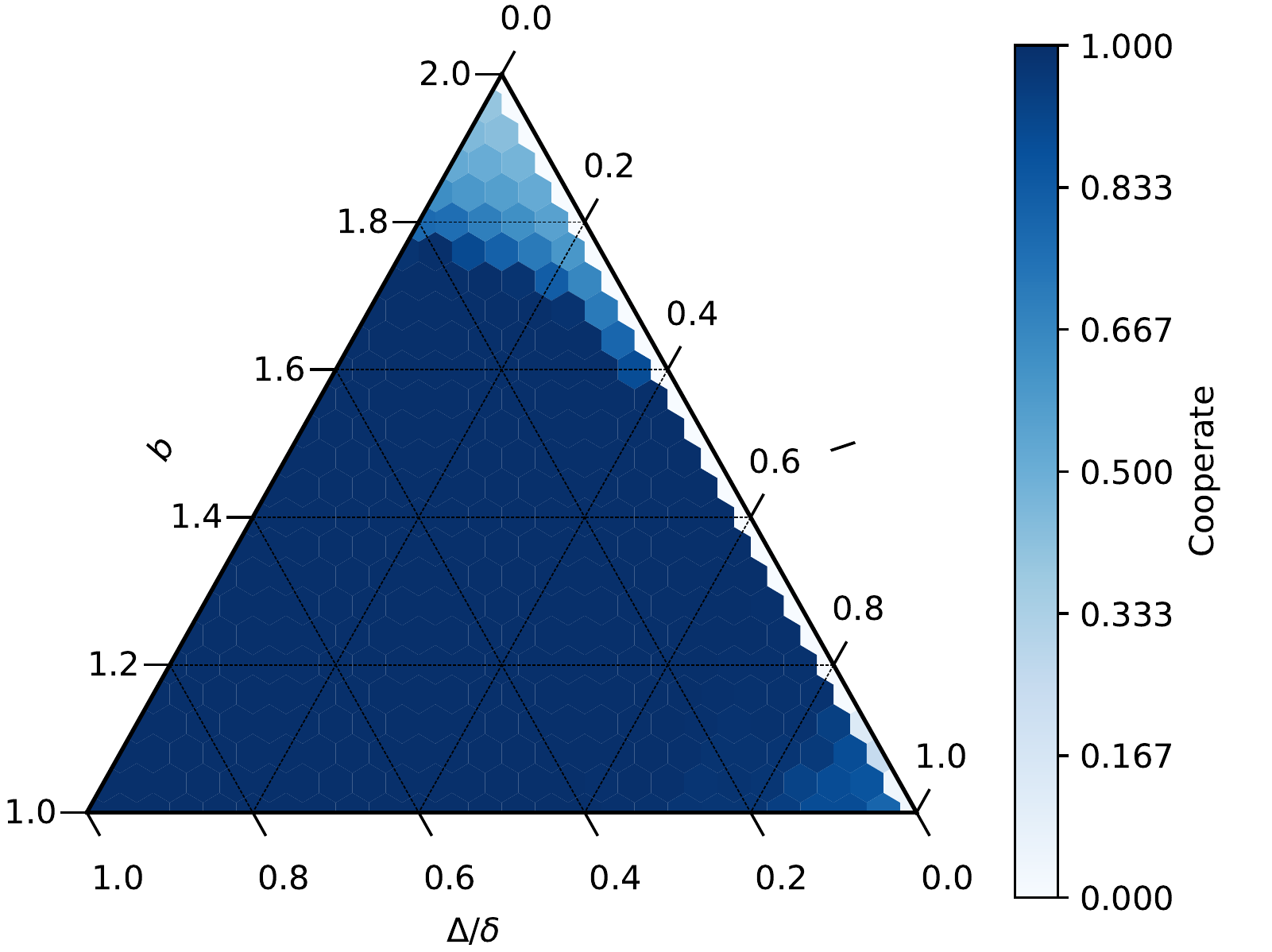, width=0.325\linewidth}}
    {\epsfig{file=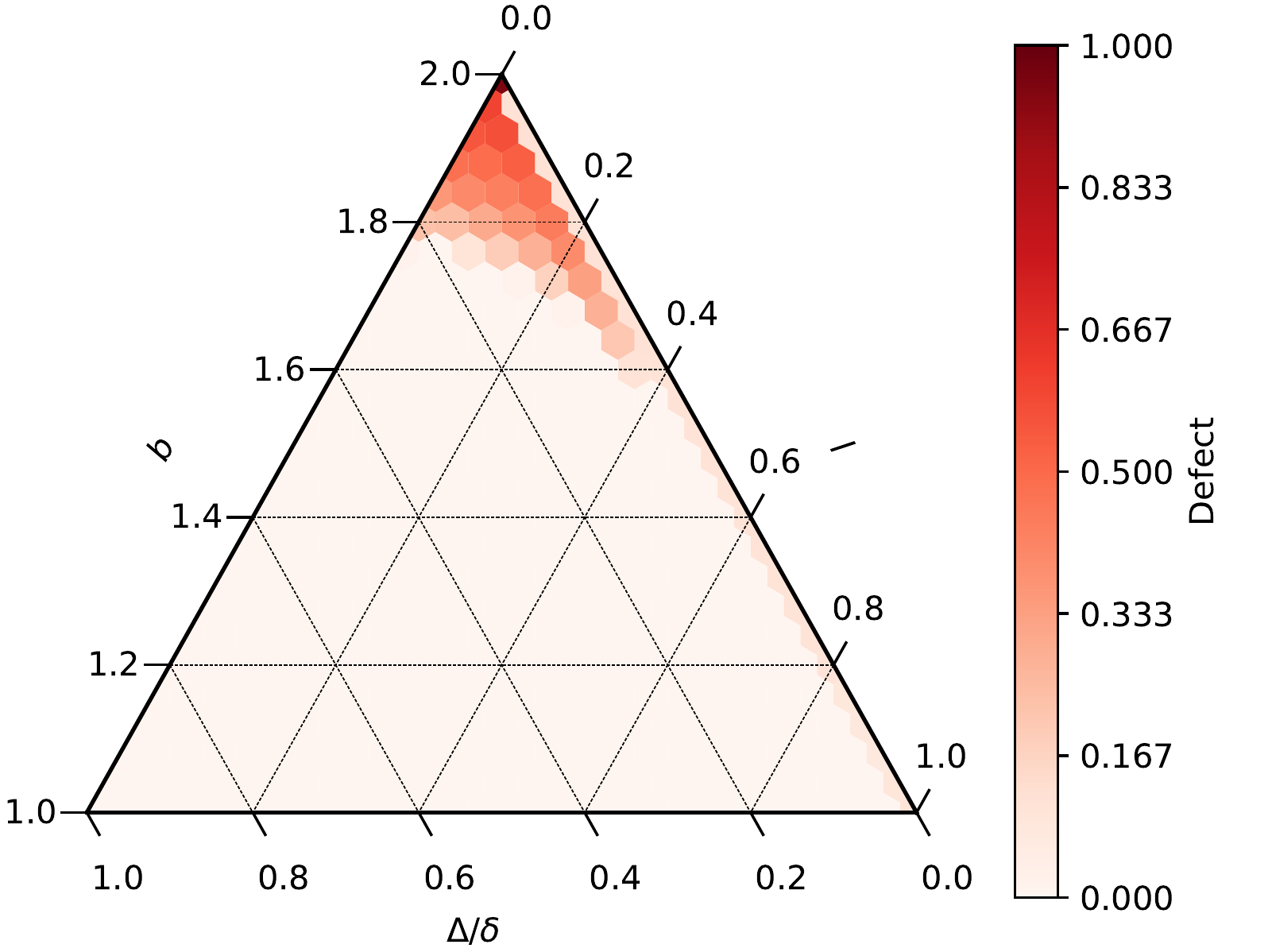, width=0.325\linewidth}}
    {\epsfig{file=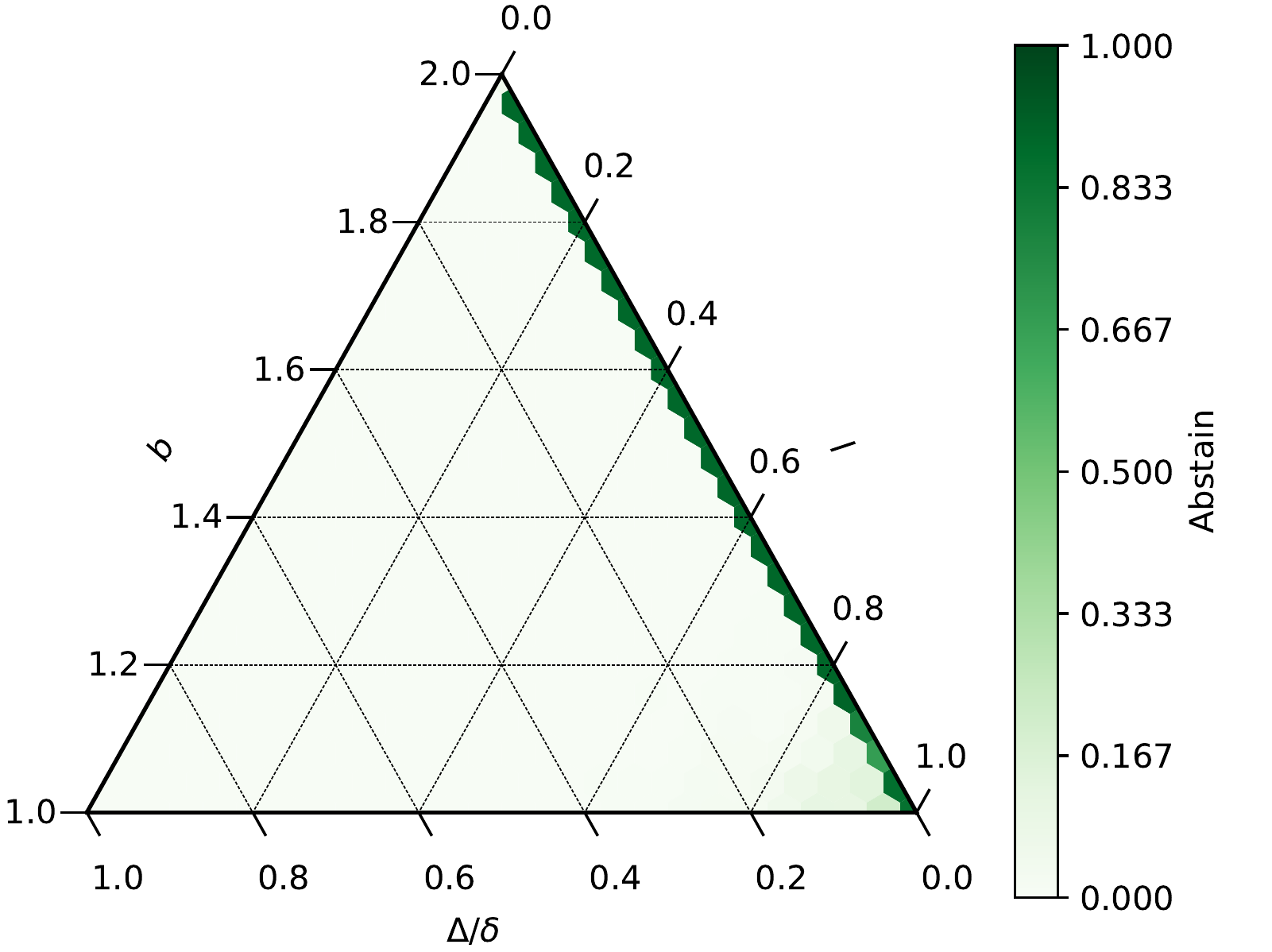, width=0.325\linewidth}}
    \caption{
        Ternary diagrams of different values of $b$, $l$ and $\Delta/\delta$
        for $\delta=0.8$.
    }
    \label{fig:ternary}
\end{figure}

As shown in Figure~\ref{fig:ternary}, cooperation is the dominant strategy in
the majority of cases. Note that in the traditional case, with an unweighted
and static network, i.e., $\Delta/\delta=0.0$, abstainers dominate in all
scenarios illustrated in this ternary diagram. In addition, it is also possible
to observe that certain combinations of $l$, $b$ and $\Delta/\delta$ guarantee
higher levels of cooperation. In these scenarios, cooperators are protected by
abstainers against exploitation from defectors.

Another observation is that defectors are attacked more efficiently by
abstainers as we increase the loner’s payoff ($l$). Simulations reveal that,
for any scenario, if the loner’s payoff is greater than $0.7$ ($l > 0.7$),
defectors have no chance of surviving.

However, the drawback of increasing the value of $l$ is that it makes it difficult
for cooperators to dominate abstainers, which might produce a quasi-stable
population of cooperators and abstainers. It is noteworthy that it is a
counter-intuitive result from the COPD game, since the loner’s payoff is always
less than the reward for mutual cooperation (i.e., $L < R$), even for extremely
high values of $L$. This scenario (population of cooperators and abstainers)
should always lead cooperators to quickly dominate the environment.

In fact, it is still expected that, in the COPD game, cooperators dominate
abstainers, but depending on the value of the loner’s payoff, or the amount of
abstainers in the population at this stage, it might take several Monte Carlo
steps to reach a stable state, which is usually a state of cooperation fully
dominating the population.

An interesting behaviour is noticed when $l=[0.45, 0.55]$ and $b > 1.8$. In
this scenario, abstainers quickly dominate the population, making a clear
division between two states: before this range (defectors hardly die off) and
after this range (defectors hardly survive). In this way, a loner’s payoff value
greater than $0.55$ ($l > 0.55$) is usually the best choice to promote
cooperation.  This result is probably related to the difference between the
possible utilities for each type of interaction, which still needs further
investigation in future.

Although the combinations shown in Figure~\ref{fig:ternary} for higher values
of b ($b>1.8$) are just a small subset of an infinite number of possible
values, it is clearly shown that a reasonable fraction of cooperators can
survive even in an extremely adverse situation where the advantage of defecting
is very high.  Indeed, our results show that some combinations of high values
of $l$ and $\delta$, such as for $\delta=0.8$ and $l=0.7$, can further improve the
levels of cooperation, allowing for the full dominance of cooperation.

\section{Investigating the Robustness of Cooperation in a Biased Environment}
\label{sec:results3}

The previous experiments revealed that the presence of abstainers together with
simple coevolutionary rules (i.e., the COPD game) act as a powerful mechanism to
avoid the spread of defectors, which also allows the dominance of cooperation
in a wide range of scenarios.

However, the distribution of the strategies in the initial population used in
all of the previous experiments was uniform. That is, we have explored cases in which
the initial population contained a balanced amount of cooperators, defectors and
abstainers. Thus, in order to explore the robustness of these outcomes in
regard to the initial amount of abstainers in the population, we now aim to
investigate how many abstainers would be necessary to guarantee robust
cooperation.

\begin{figure}[p]
    \centering
    {\epsfig{file=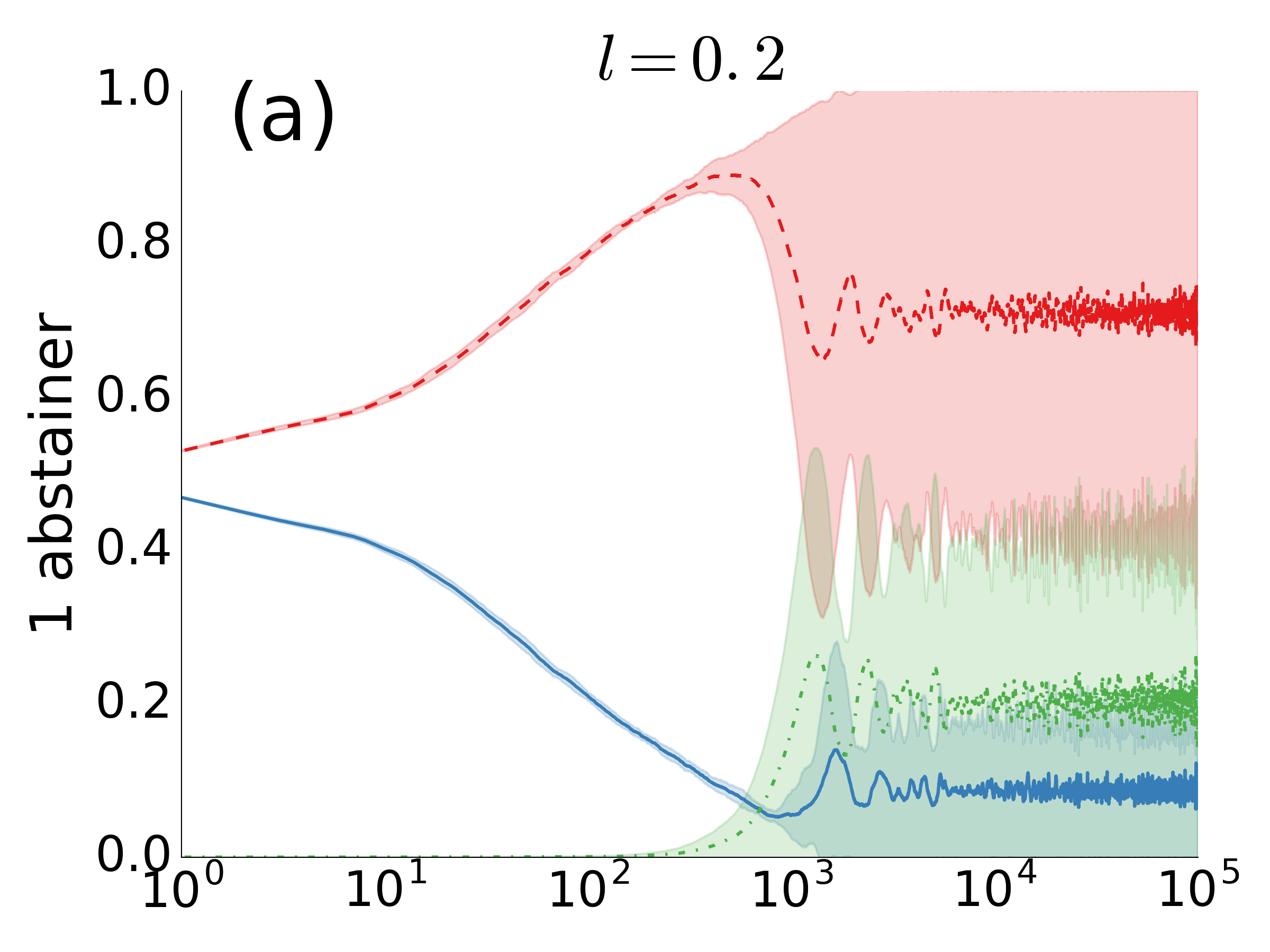, width=0.325\linewidth}}
    {\epsfig{file=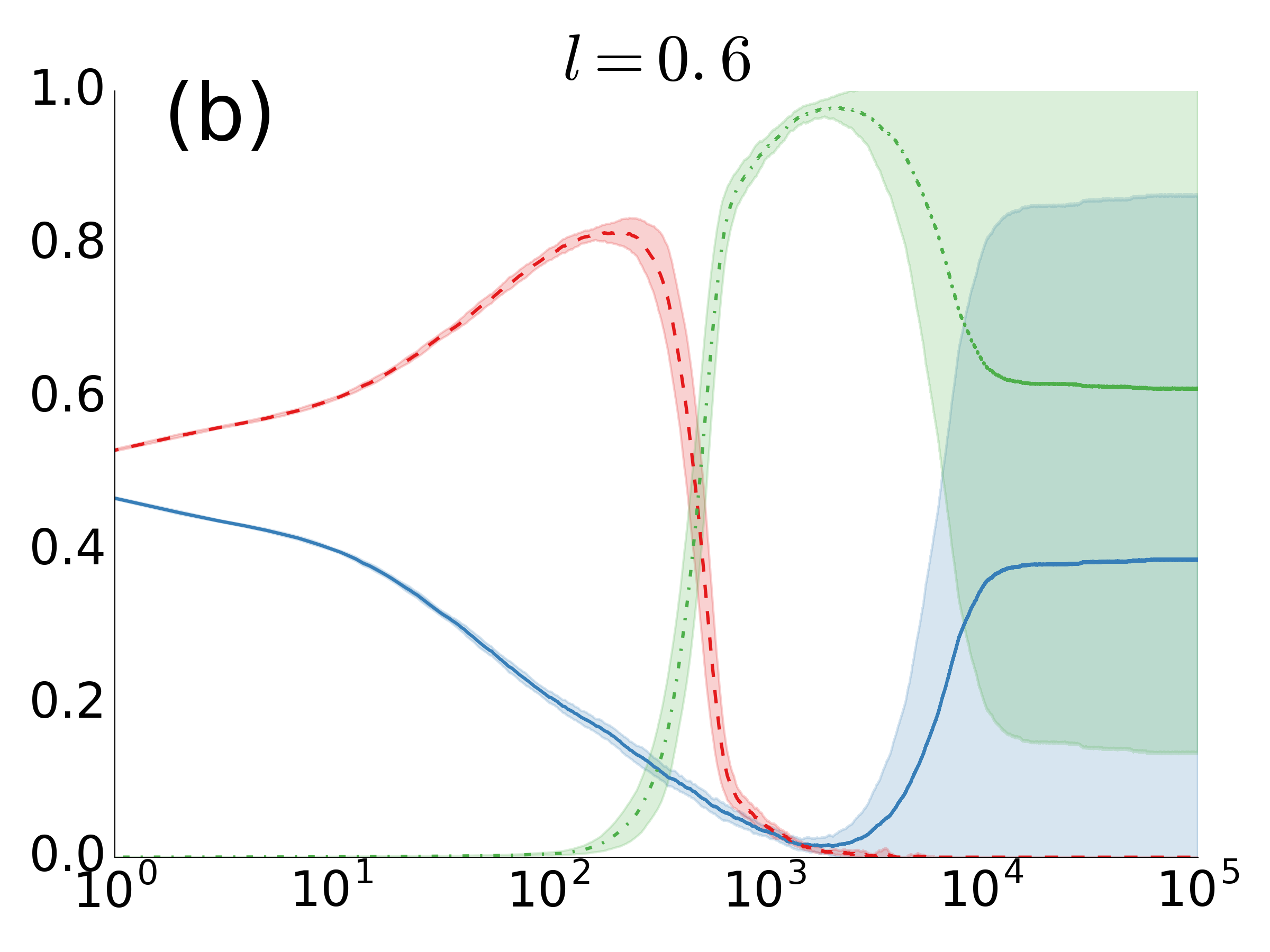, width=0.325\linewidth}}
    {\epsfig{file=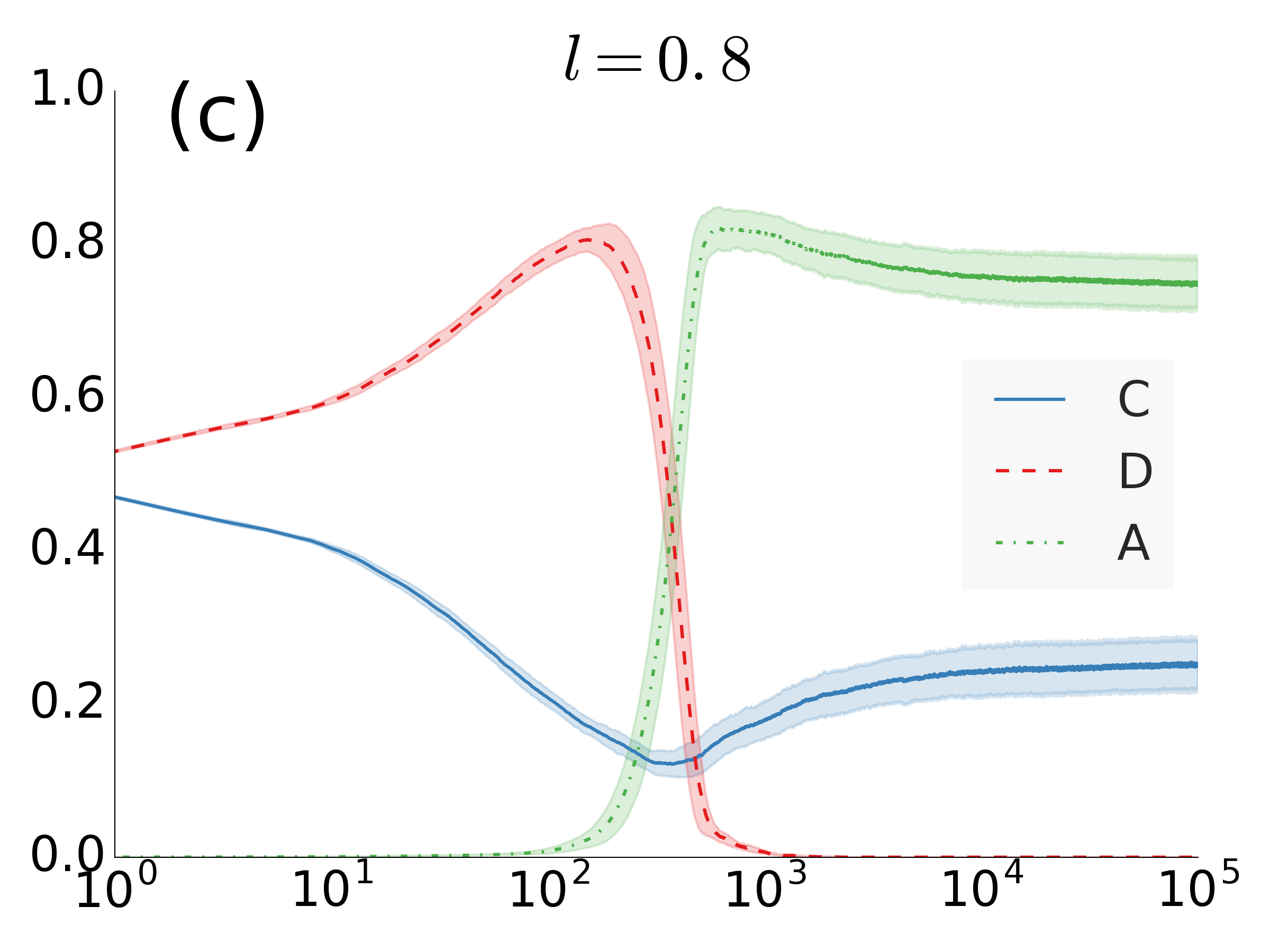, width=0.325\linewidth}}

    {\epsfig{file=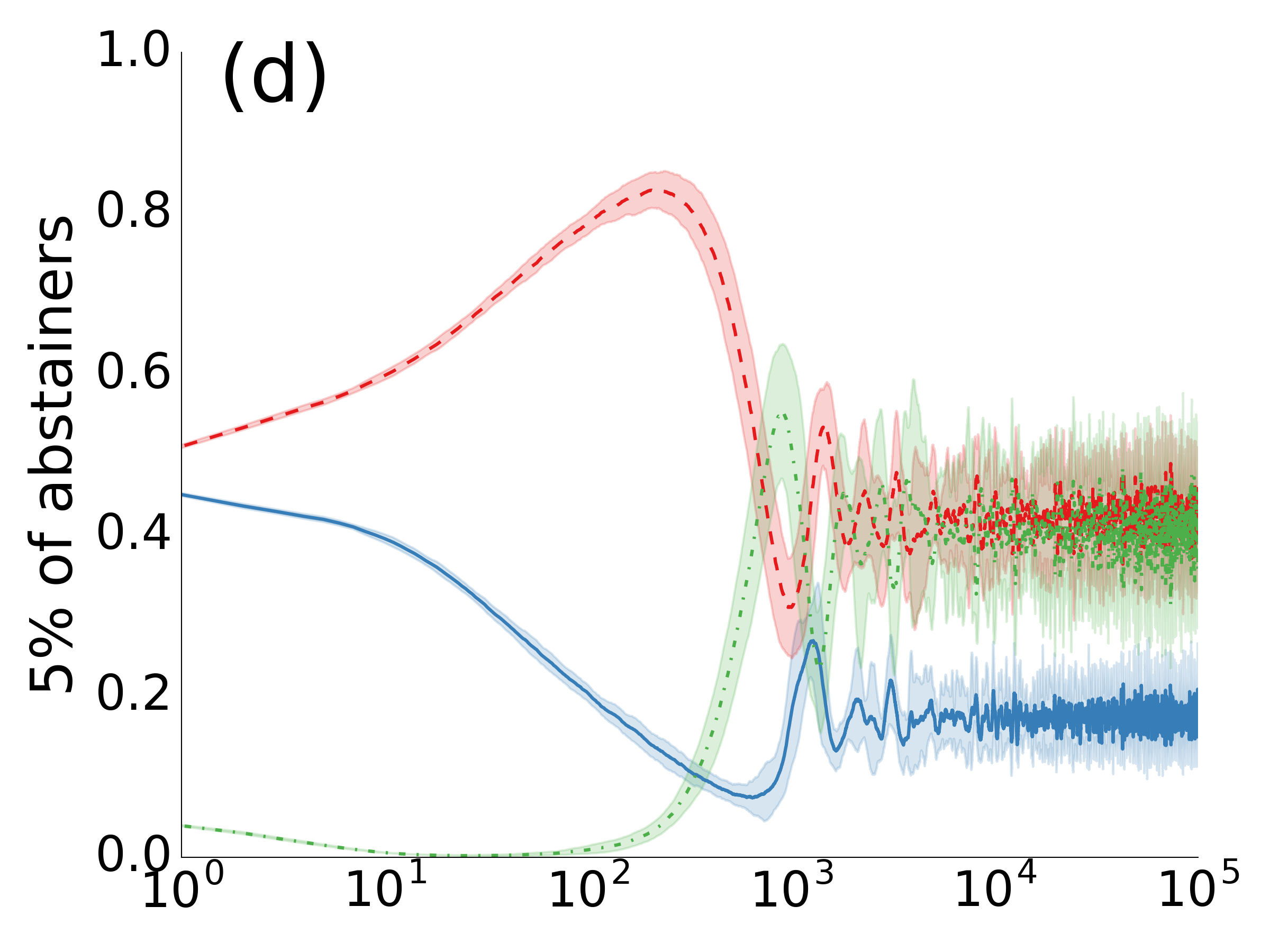, width=0.325\linewidth}}
    {\epsfig{file=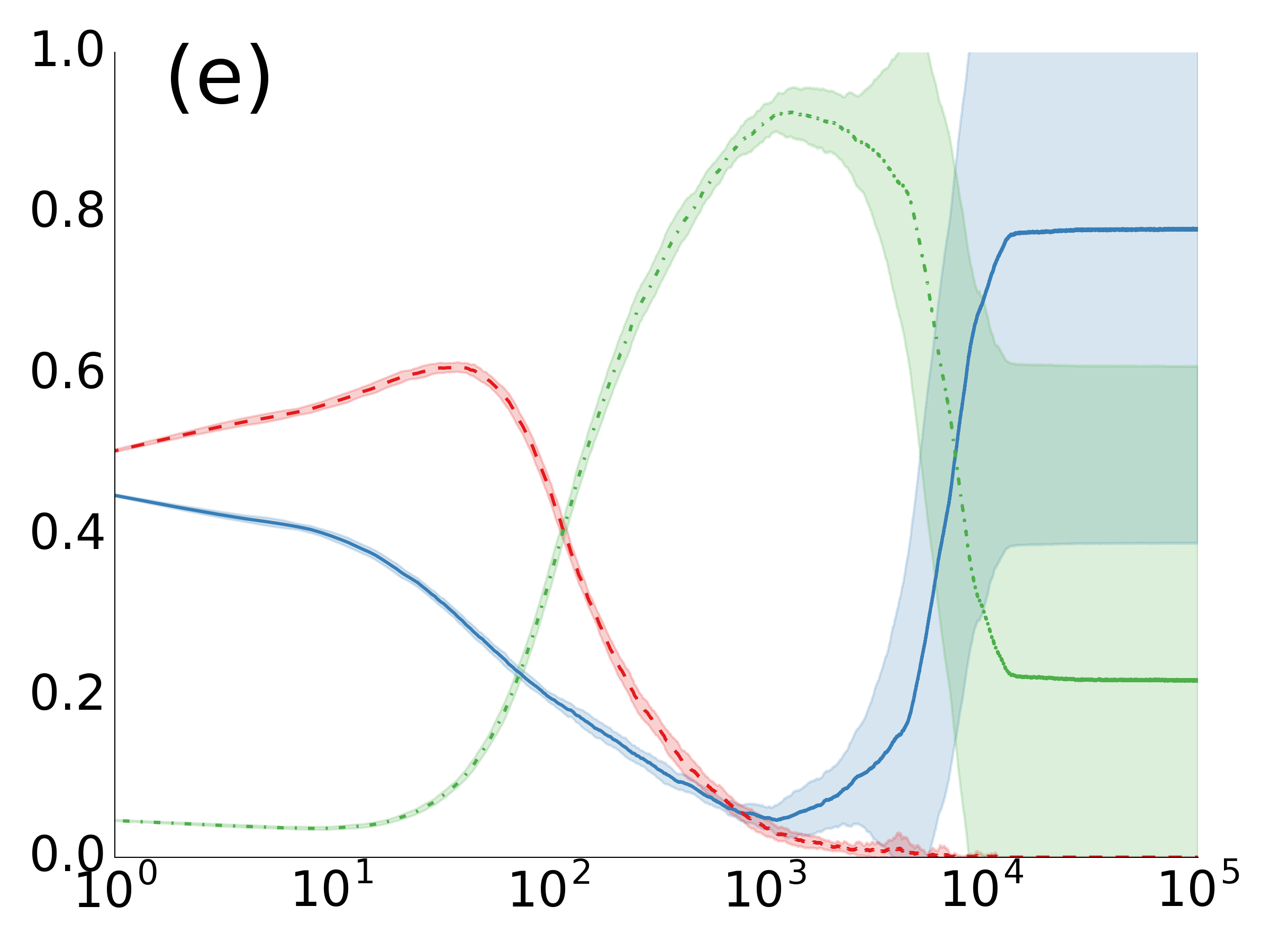, width=0.325\linewidth}}
    {\epsfig{file=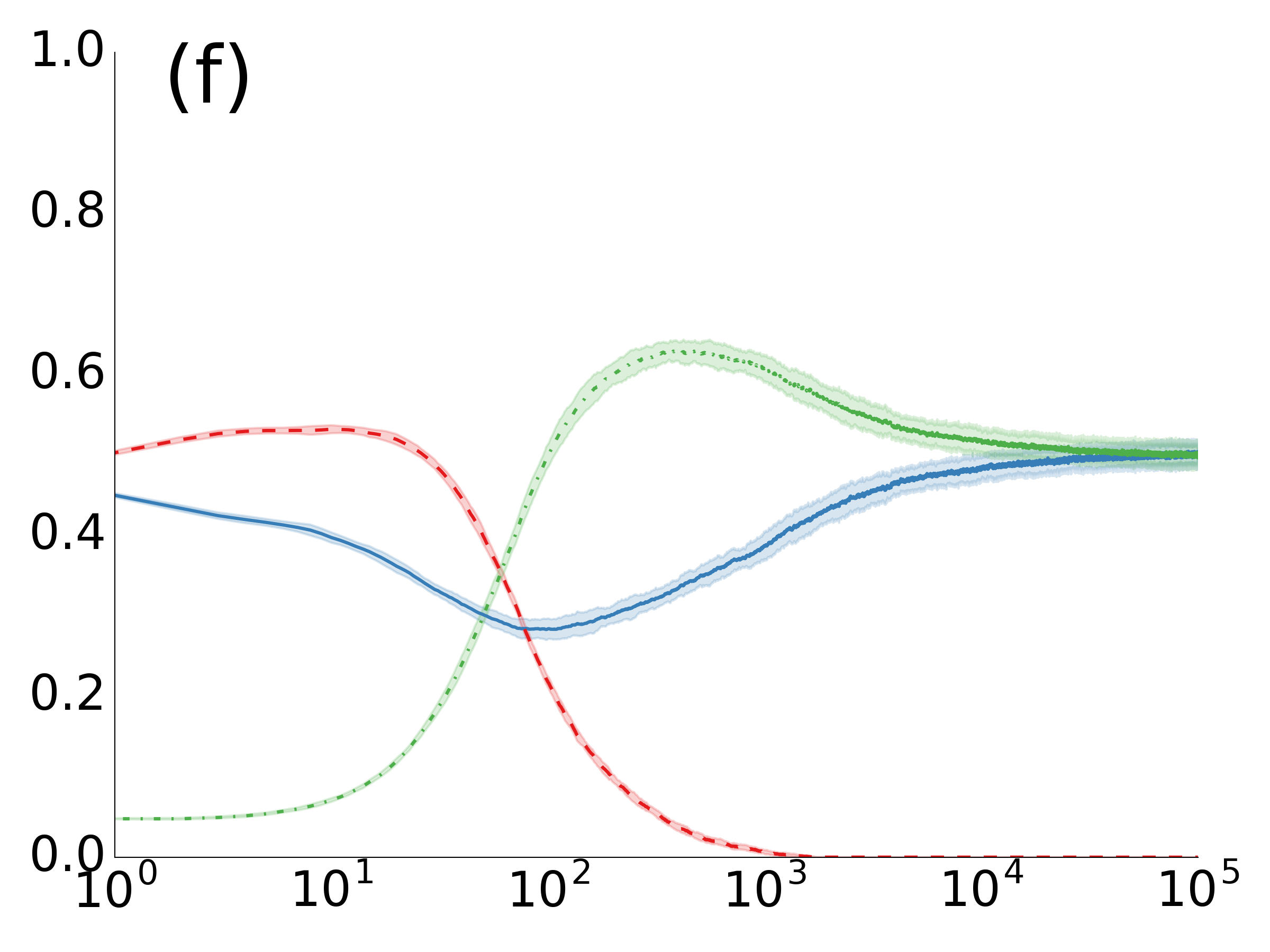, width=0.325\linewidth}}

    {\epsfig{file=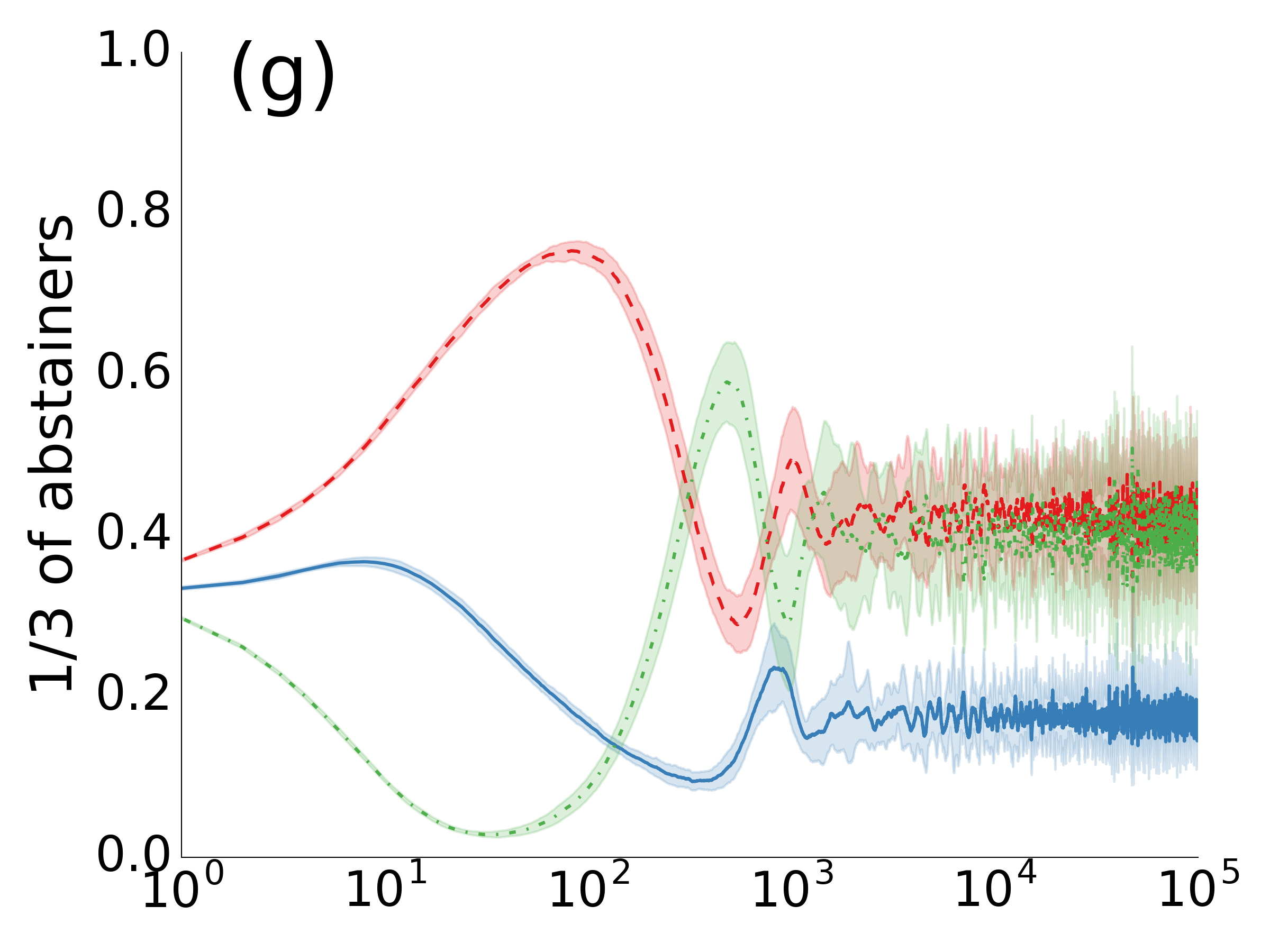, width=0.325\linewidth}}
    {\epsfig{file=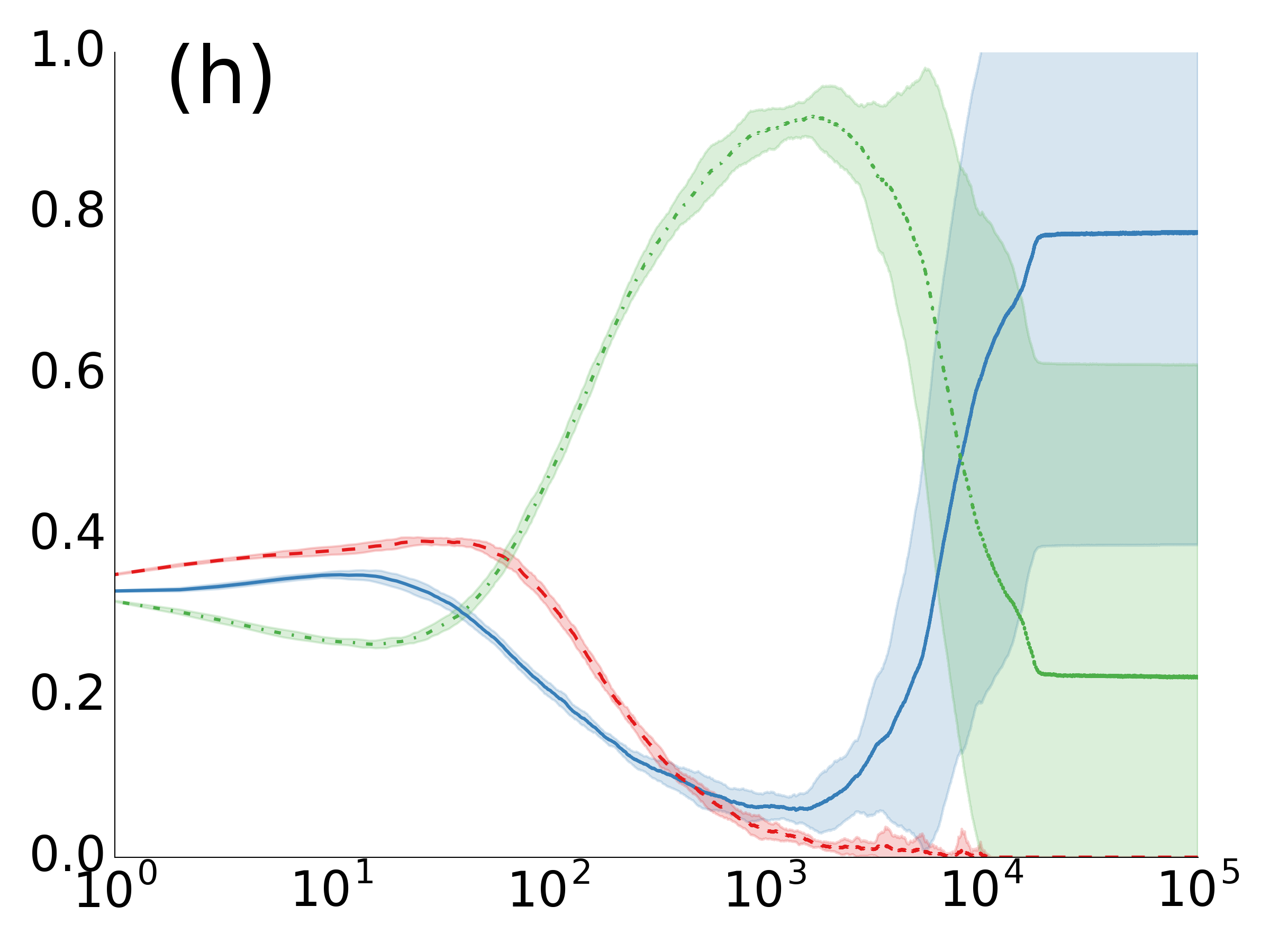, width=0.325\linewidth}}
    {\epsfig{file=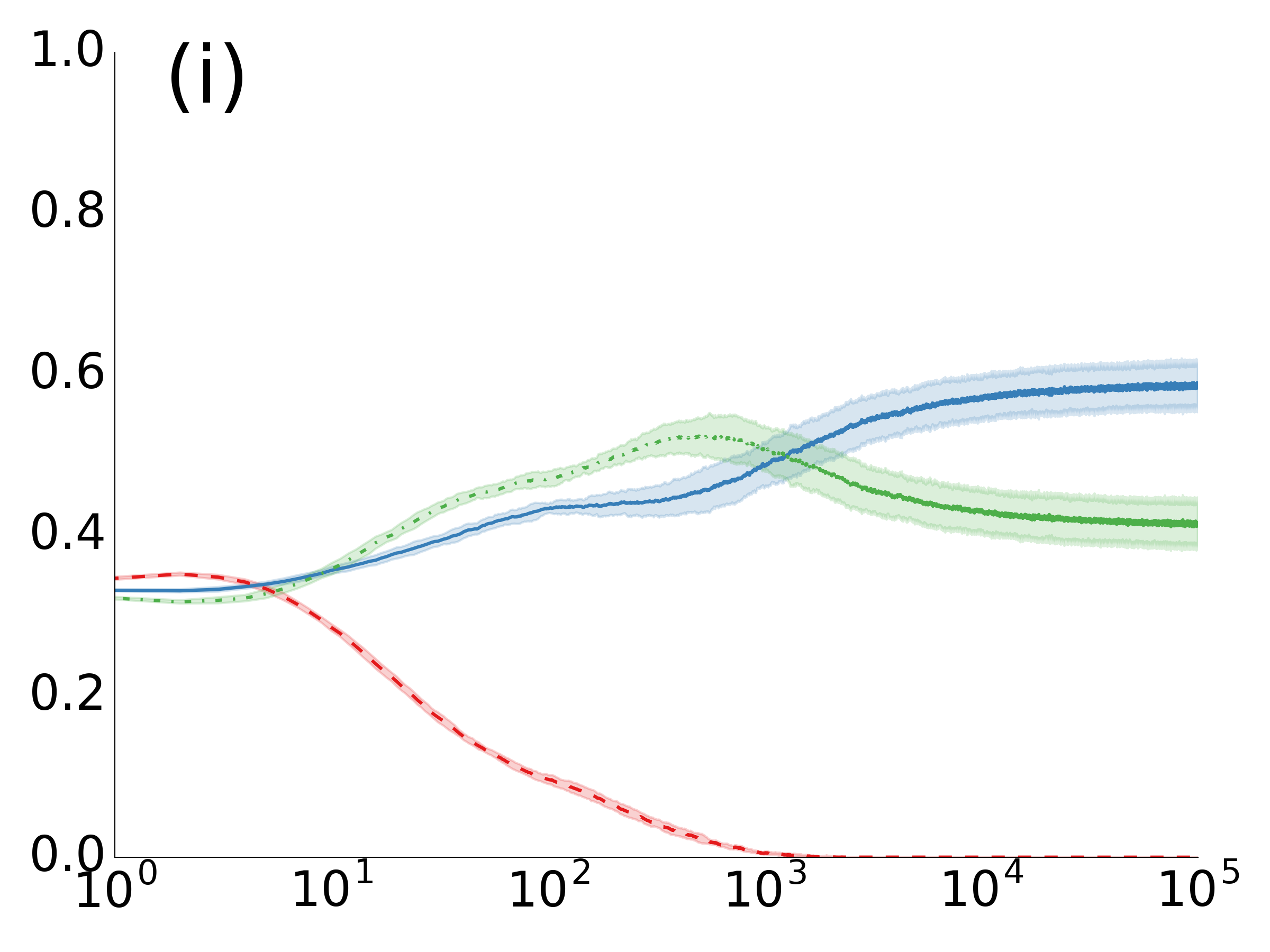, width=0.325\linewidth}}

    {\epsfig{file=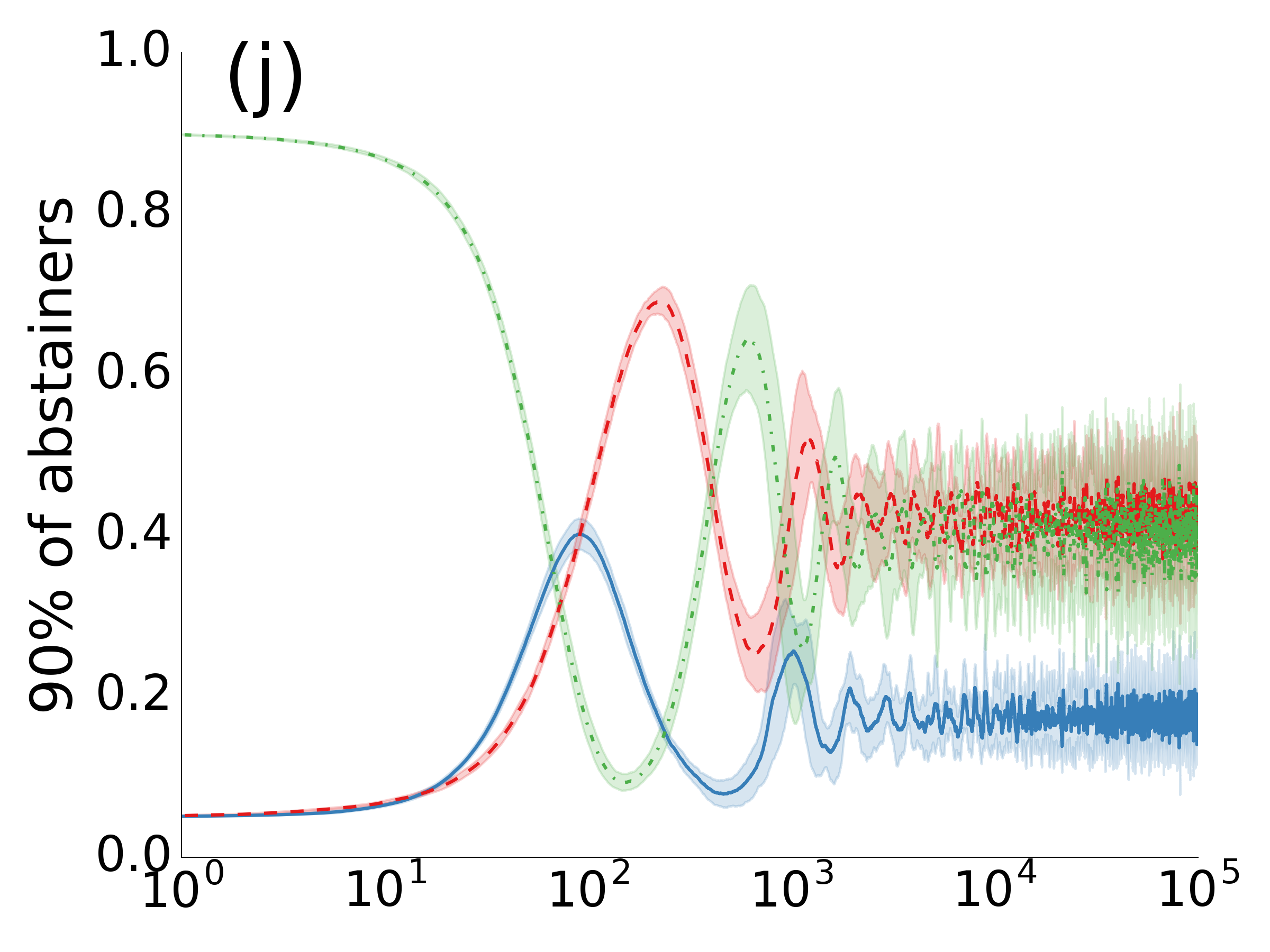, width=0.325\linewidth}}
    {\epsfig{file=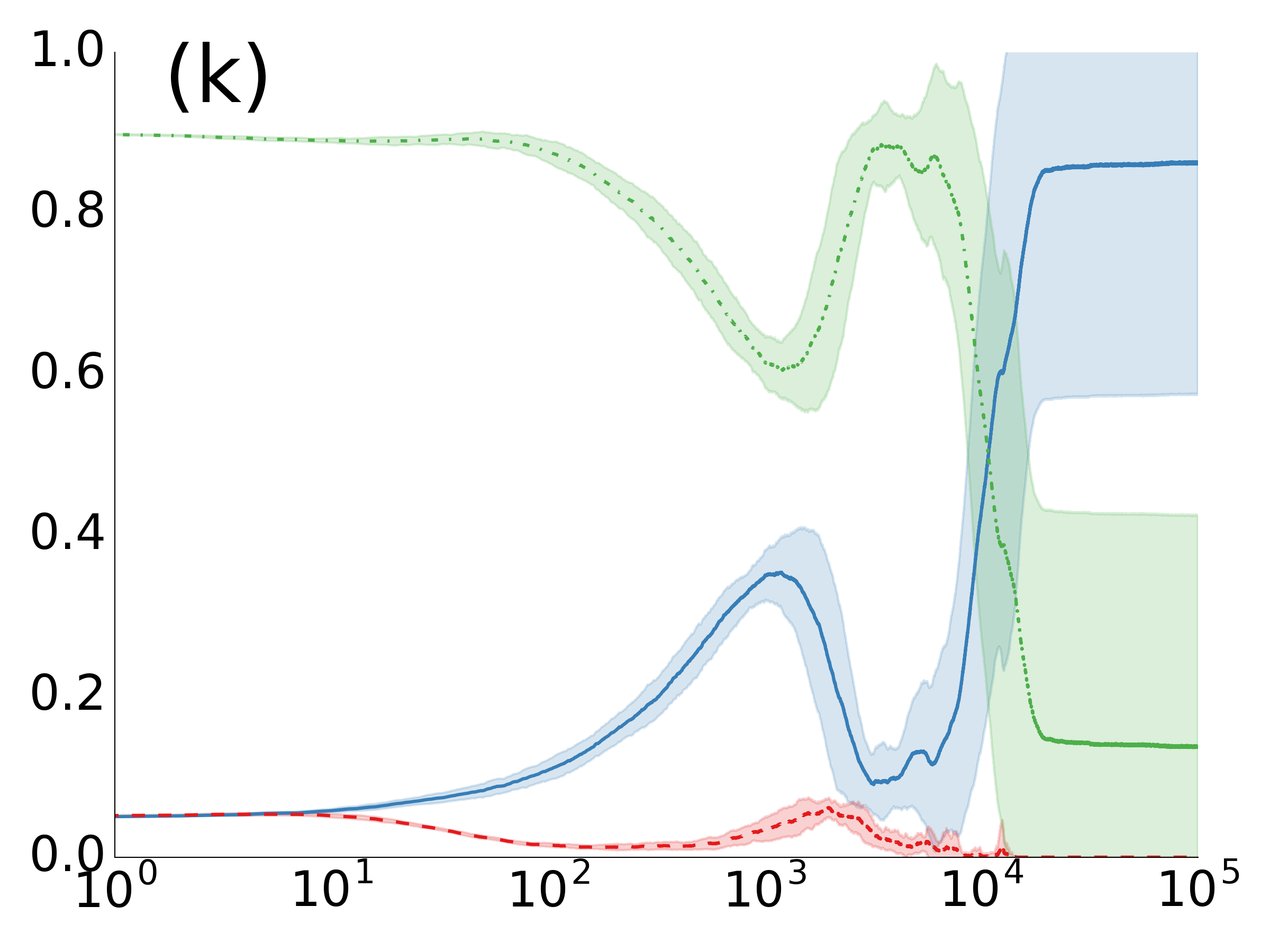, width=0.325\linewidth}}
    {\epsfig{file=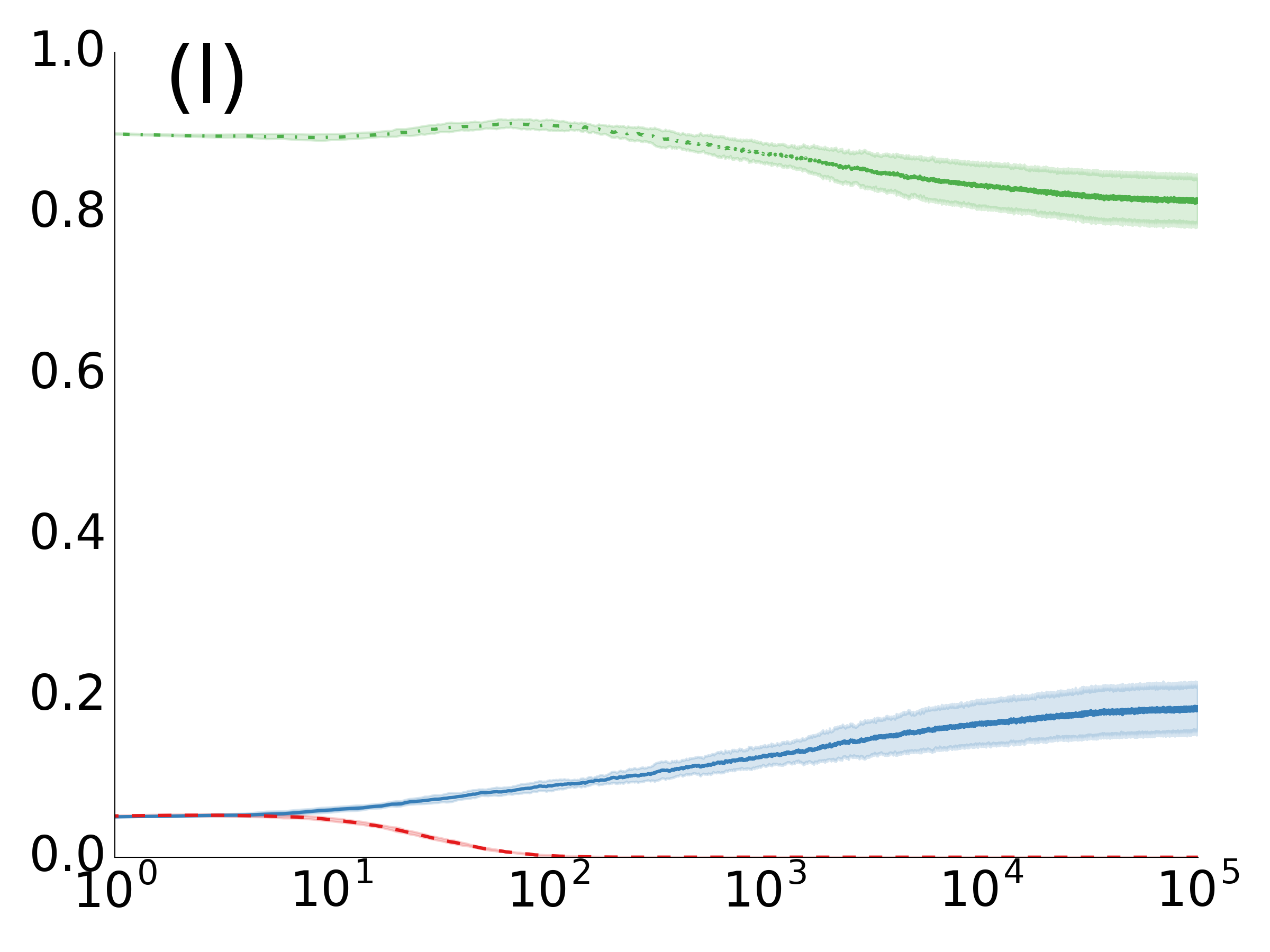, width=0.325\linewidth}}

    \caption{
        Time course of each strategy for $b=1.9$, $\Delta=0.72$,
        $\delta=0.8$ and different values of $l$ (from left to right,
        $l=\{0.2,\ 0.6,\ 0.8\}$). The same settings are also tested
        on populations seeded with different amount of abstainers
        (i.e, from top to bottom: 1 abstainer, 5\% of the population,
        1/3 of the population, 90\% of the population).
    }
    \label{fig:unbalanced}
\end{figure}

Figure~\ref{fig:unbalanced} features the fraction of each strategy in the
population (i.e., cooperators, defectors and abstainers) over time for fixed
values of $b=1.9$, $\Delta=0.72$ and $\delta=0.8$. In this experiment, several
independent simulations were performed, in which the loner’s payoff ($l$) and the
number of abstainers in the initial population were varied from $0.0$ to $1.0$
and from $0.1\%$ to $99.9\%$, respectively. Other special cases were also
analyzed, such as placing only one abstainer into a balanced population of
cooperators and defectors, and placing only one defector and one cooperator in
a population of abstainers. For the sake of simplicity, we report only the
values of $l=\{0.2,\ 0.6,\ 0.8\}$ for an initial population with one abstainer
and with $5\%$, $33\%$ and $90\%$ abstainers, which are representative of the
outcomes at other values also.

Note that, for all these simulations, the initial population of cooperators and
defectors remained in balance. For instance, an initial population with $50\%$
of abstainers, will consequently have $25\%$ of cooperators and $25\%$ of
defectors.

Experiments reveal that the COPD game is actually extremely robust to radical
changes in the initial population of abstainers. It has been shown that if the
loner’s payoff is greater than $0.55$ ($l>0.55$), then one abstainer might alone
be enough to protect cooperators from the invasion of defectors (see Figures
\ref{fig:unbalanced}a, \ref{fig:unbalanced}b and \ref{fig:unbalanced}c).
However, this outcome is only possible if the single abstainer is in the middle
of a big cluster of defectors.

This outcome can happen because the payoff obtained by the abstainers is always
greater than the one obtained by pairs of defectors (i.e., $L<P$). Thus, in a
cluster of defectors, abstention is always the best choice. However, as this single
abstainer reduces the population of defectors, which consequently increases the
population of abstainers and cooperators in the population, defection may start
to be a good option again due to the increase of cooperators. Therefore, the
exploitation of defectors by abstainers must be as fast as possible, otherwise,
they might not be able to effectively attack the population of defectors. In
this scenario, the loner’s payoff is the key parameter to control the speed in
which abstainers invade defectors. This explains why a single abstainer
is usually not enough to avoid the dominance of defectors when $l<0.55$.

In this way, as the loner’s payoff is the only parameter that directly affects
the evolutionary dynamics of the abstainers, intuition might lead one to expect
to see a clear and perhaps linear relationship between the loner’s payoff and
the initial number of abstainers in the population. That is, given the same set
of parameters, increasing the initial population of abstainers or the loner’s
payoff ($l$) would probably make it easier for abstainers to increase or even
dominate the population. Despite the fact that it might be true for high values
of the loner’s payoff (i.e., $l\ge0.8$, as observed in Figure~\ref{fig:unbalanced})
is not applicable to other scenarios. Actually, as it is also shown in
Figure~\ref{fig:unbalanced}, if the loner’s payoff is less than $0.55$,
changing the initial population of abstainers does not change the outcome at
all. When $0.55 \le l < 0.8$, a huge initial population of abstainers can
actually promote cooperation best.

As discussed in Section~\ref{sec:varyall}, populations of cooperators and
abstainers tend to converge to cooperation. In this way, the scenario showed in
Figure~\ref{fig:unbalanced} for $l=0.8$  will probably end up with cooperators
dominating the population, but as the loner’s payoff is close to the reward for
mutual cooperation, the case in Figure~\ref{fig:unbalanced}i will converge
faster than the one showed in Figure\ref{fig:unbalanced}l.

Another very counter-intuitive behaviour occurs in the range $l=[0.45, 0.55]$
(this range may shift a little bit depending on the value of $b$), where the
outcome is usually of abstainers quickly dominating the population
(Sect.\ref{sec:varyall}). In this scenario, we would expect that changes in the
initial population of abstainers would at least change the speed in which the
abstainers fixate in the population. That is, a huge initial population of
abstainers would probably converge quickly. However, it was observed that the
convergence speed is almost the same regardless of the size of the initial
population of abstainers.

In summary, results show that an initial population with $5\%$ of abstainers
is usually enough to make reasonable changes in the outcome, increasing the
chances of cooperators surviving or dominating the population.

\section{Conclusions and Future Work}
\label{sec:conclusion}

This paper studies the impact of a simple coevolutionary model in which not
only the agents’ strategies but also the network evolves over time. The model
consists of placing agents playing the Optional Prisoner’s Dilemma game in a
dynamic spatial environment, which in turn, defines the Coevolutionary Optional
Prisoner’s Dilemma (COPD) game.

In summary, based on the results of several Monte Carlo simulations, it was
shown that the COPD game allows for the emergence of cooperation in a wider
range of scenarios than the Coevolutionary Prisoner’s Dilemma (CPD) game (i.e.,
the same coevolutionary model in populations which do not have the option to
abstain from playing the game). Results also showed that COPD performs much
better than the traditional version of these games (i.e., the Prisoner’s
Dilemma (PD) and the Optional Prisoner’s Dilemma (OPD) games) where only the
strategies evolve over time in a static and unweighted network.
Moreover, we observed that the COPD game is actually able to reproduce outcomes
similar to other games by setting the parameters as follows:
\begin{itemize}
    \item CPD: $l=0$.
    \item OPD: $\Delta=0$ (or $\delta=0$).
    \item PD: $l=0$ and $\Delta=0$ (or $\delta=0$).
\end{itemize}

Also, it was possible to observe that abstention acts as an important
mechanism to avoid the dominance of defectors. For instance, in adverse
scenarios such as when the defector’s payoff is very high (i.e., $b>1.7$), for both
PD and CPD games, defectors spread quickly and dominated the environment. On
the other hand, when abstainers were present (COPD game), cooperation was able
to survive and even dominate.

Furthermore, simulations showed that defectors die off when the loner’s payoff is
greater than $0.7$ ($l>0.7$). However, it was observed that increasing the loner’s
payoff makes it difficult for cooperators to dominate abstainers, which is a
counter-intuitive result, since the loner’s payoff is always less than the
reward for mutual cooperation (i.e., $L < R$), this scenario should always lead
cooperators to dominance very quickly. In this scenario, cooperation is still
the dominant strategy in most cases, but it might require several Monte Carlo
steps to reach a stable state.

Results revealed that the COPD game also allows scenarios of cyclic dominance
between the three strategies (i.e., cooperation, defection and abstention),
indicating that, for some parameter settings, the COPD game is intransitive.
That is, the population remains balanced in such a way that cooperators invade
abstainers, abstainers invade defectors and defectors invade cooperators,
closing a cycle.

We also explored the robustness of these outcomes in regard to the initial
amount of abstainers in the population (biased population). In summary, it was
shown that, in some of the scenarios, even one abstainer might alone be enough
to protect cooperators from the invasion of defectors, which in turn increases
the chances of cooperators surviving or dominating the population.

Although recent research has considered coevolving game strategy (with optional
games) and link weights \cite{Cardinot2016ecta}, this work presents a more complete
analysis. We conclude that the combination of both of these trends in
evolutionary game theory may shed additional light on gaining an in-depth
understanding of the emergence of cooperative behaviour in real-world
scenarios.

Future work will consider the exploration of different topologies and the
influence of a wider range of scenarios, where, for example, agents could
rewire their links, which, in turn, adds another level of complexity to the
model. Future work will also involve applying our studies and results to
realistic scenarios, such as social networks and real biological networks.

\bigskip
\subsubsection*{Acknowledgments. }
This work was supported by the National Council for Scientific and Technological Development (CNPq-Brazil).

\bibliographystyle{splncs03}
\bibliography{refs}

\end{document}